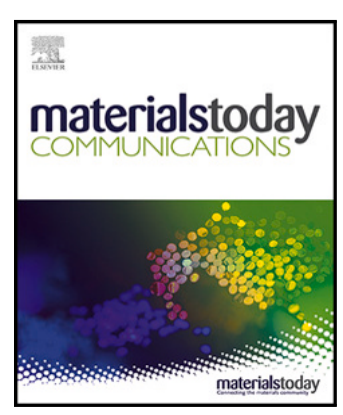

# Interpretation of field emission current–voltage data: Background theory and detailed simulation testing of a user-friendly webtool

Mohammad M. Allaham [a,b], Richard G. Forbes [c,*,1], Alexandr Knápek [a,*,2], Dinara Sobola [d,e], Daniel Burda [a,e], Petr Sedlák [e], Marwan S. Mousa [f]

[a] Institute of Scientific Instruments of Czech Academy of Sciences, Královopolská 147, 612 64 Brno, Czech Republic
[b] Central European Institute of Technology, Brno University of Technology, Purkyňova 123, 612 00 Brno, Czech Republic
[c] Advanced Technology Institute and Department of Electrical and Electronic Eng., University of Surrey, Guildford, Surrey GU2 7XH, UK
[d] Institute of Physics of Materials, Academy of Sciences CR, Zizkova 22, 616 62 Brno, Czech Republic
[e] Department of Physics, Faculty of Electrical Engineering and Communication, Brno University of Technology, Technická 2848/8, 616 00 Brno, Czech Republic
[f] Department of Physics, Mu'tah University, Al-Karak 61710, Jordan

## ARTICLE INFO

## ABSTRACT

In field electron emission (FE) studies, interpretation of measured current–voltage characteristics and extraction of emitter characterization parameters are usually carried out within the framework of "smooth planar metal-like emitter (SPME) methodology", using a data-analysis plot. This methodology was originally introduced in the 1920s. Three main data-plot types now exist: Millikan–Lauritsen (ML) plots, Fowler–Nordheim (FN) plots, and Murphy–Good (MG) plots. ML plots were commonly used in early FE studies, but most modern analysis uses FN plots. MG plots are a recent introduction.

Theoretically, it is now known that ML and FN plots are predicted to be slightly curved in SPME methodology, but a Murphy–Good plot will be very nearly straight. Hence (because 1956 Murphy–Good emission theory is "better physics" than 1928 Fowler–Nordheim emission theory as corrected in 1929), expectation is that parameter extraction using a MG plot will be more precise than extraction using either ML plots or FN plots.

In technological FE studies, current–voltage characteristics are often converted into other forms. Thus, measured voltage may be converted to (apparent) macroscopic field, and/or current values may be converted to macroscopic current densities. Thus, four data-input forms can be found in the context of analysing FE current–voltage results.

It is also the case that over-simplified models of measurement-system behaviour are very widely assumed, and the question of whether simple use of a data-analysis plot is a valid data-interpretation procedure for the particular system under investigation has often been neglected. Past published studies on field emitter materials development appear to contain a high incidence of spurious values for the emitter characterization parameter "characteristic field enhancement factor". A procedure (the so-called "Orthodoxy Test") was described in 2013 that allows a validity check on measurement-system behaviour, and found that around 40% of a small sample of results tested were spuriously high, but has had limited uptake so far.

To assist with FE current–voltage data interpretation and validity checks, a simple user-friendly webtool has been under design by the lead author. The webtool needs as user input some system specification data and some "range-limits" data from any of the three forms of data-analysis plot, using any of the four data-input variations. The webtool then applies the Orthodoxy Test, and—if the Test is passed—calculates values of relevant emitter characterization parameters.

The present study reports the following: (1) systematic tests of the webtool functionality, using simulated input data prepared using Extended Murphy–Good field electron emission theory; and (2) systematic comparisons of the three different data-plot types, again using simulated input data, in respect of the accuracy with which extracted characterization parameter values match the simulation input values. The paper is introduced by a thorough summary review of the theory on which modern SPME-based current–voltage data-analysis procedures are based. The need in principle to move on (in due course) to data-analysis procedures based on curved-emitter emission theory is noted.

* Corresponding authors.
*E-mail addresses:* r.forbes@trinity.cantab.net (R.G. Forbes), knapek@isibrno.cz (A. Knápek).

[1] Theory.

[2] Simulation results.

An important result is to confirm (by simulations) that, particularly in respect of the extraction of formal emission areas, the performance of the Murphy–Good plot is noticeably better than the performances of Fowler–Nordheim and Millikan–Lauritsen plots. This result is important for field electron emission science because it is now known that differences as between different theories of field electron emission often affect the formal emission area.

## 1. Introduction

An important technique for generating electron beams from a metallic electron source (called here the *emitter*), is the well-known process of *cold field electron emission (CFE)* [1–6]. Emitter electrons from a population in local thermodynamic equilibrium, at or near room temperature, are extracted from electron-energy states close to the Fermi level (FL), under the influence of a negative surface electrostatic field of high absolute magnitude $F_{\mathrm{L}}$, typically a few V/nm.

There are several different mathematical theories of this process (based on different physical assumptions), but they all generate formulae for the *local emission current density* (LECD) $J_{\mathrm{L}}$ in terms of the local work function $\phi_{\mathrm{L}}$, the local field magnitude $F_{\mathrm{L}}$ and (in some cases) other emitter characterization parameters.

Although more advanced emission theories do now exist, emission aspects of most experimental CFE work are still interpreted within the framework of a basic underlying theoretical model originally introduced into CFE theory in the 1920s. This disregards the existence of atoms and the possible role of surface-atom atomic orbitals in the emission process, and assumes that the emitter can be treated as a Sommerfeld-type free-electron conductor with a smooth, planar, structureless surface of large lateral extent. This methodology has been called [7] *smooth planar metal-like emitter (SPME) methodology*. Mathematical emission theories derived in the context of SPME methodology are called here "smooth-surface planar field electron emission (FE) theories".

SPME methodology can in practice be applied to needle-shaped and post-shaped emitters that are "not too sharp" by making the so-called *planar emission approximation*, and carrying out an integration of the LECD over the emitter surface. The planar emission approximation involves the assumption that, at any given point "P" on the needle (etc.) surface, the LECD is given by the planar FE equation of interest, using the values of $\phi_{\mathrm{L}}$ and $F_{\mathrm{L}}$ at point P. This is an approximation because, in reality, the true value of the LECD $J_{\mathrm{L}}(\phi_{\mathrm{L}}, F_{\mathrm{L}})$ is also influenced by the local surface curvature.

Within SPME methodology, the two most popular emission theories are the (long discredited—see [8]) 1928/29 theory of Fowler and Nordheim (FN)—unfortunately now widely re-introduced in a simplified form, often with a recent new form of error, and the 1956 theory of Murphy and Good (MG) [9], which corrected significant errors found in the 1920s work. The FN theory is described as the 1928/29 theory because there is a large calculational error in Eq. (22) of the 1928 paper [10] (very roughly of the order of $10^{15}$) that was rapidly found and corrected in the 1929 paper of Stern et al. [11]. The 1956 MG paper then corrected *additional* FN errors relating to the shape of the tunnelling barrier.

More recently (see [7]), one of us (RGF) has introduced so-called *Extended* Murphy–Good (EMG) theory, in which a prediction uncertainty factor is introduced into the pre-exponential of the MG local emission current density equation, in order to formally recognize various physical effects (including atomic-level effects) that 1956 MG FE theory does not include.

In past FE literature, 1956 Murphy–Good FE theory has often been called the "Fowler–Nordheim equation", and some authors continue to do this. We regard this nomenclature convention as highly confusing, and do not use it. We commend the idea that the 1956 equation should be called after its actual developers, and should be known as the "1956 Murphy–Good FE equation": its predictions for LECD are typically 100 to 500 times higher [8] than those of the 1928/29 FN FE equation.

Obviously, many modern forms of field electron emitter are not metals. All of the planar FE theories just discussed were derived in the context of (metal-based) SPME methodology. Nevertheless, it is customary practice (partly due to the lack of any easy alternative) for experimentalists interested in characterizing their emitters to use one of the two popular theories derived for metal emission.

In reality, there is some justification for using existing SPME methodologies, because what they first attempt to investigate relates to the nature of the tunnelling barrier and the electrostatics of the system geometry, and these behaviours are relatively similar for most materials, including modern non-metals. However, the existing methodologies "work adequately" only in a proportion of cases (probably around 50% or slightly more). In the remaining cases there is a high chance that extracted emitter-material characterization parameters are spurious.

The reason is not necessarily some inadequacy in emission theory. What the emission theories generate is some expression for the predicted emission current $I_{\mathrm{e}}$ in terms of some characteristic local field-magnitude $F_{\mathrm{C}}$, at some location "C" (near the emitter tip) where the LECD has a maximum value. But what is measured is the relation between the *measured current* $I_{\mathrm{m}}$ and the *measured voltage* $V_{\mathrm{m}}$: other factors may affect this.

The term *FE system* is defined to include all aspects of the experimental system that can affect the $I_{\mathrm{m}}(V_{\mathrm{m}})$ relationship, including: emitter composition, geometry and surface condition; the mechanical, geometrical and electrical arrangements in the vacuum system; all aspects of the electronic circuitry and all electronic measurement instruments; the emission physics; and ALL relevant physical processes that might be taking place (for example, the generation of field emitted vacuum space-charge, Maxwell-stress-induced reversible changes in emitter geometry, and adsorbate atom dynamics).

In general, the interpretation of FE measured current–voltage data can be a highly complex problem in *electronic engineering* that is impossible to solve exactly in the present state of research knowledge. However, for some systems, for example when the emitter and its support arrangements are good conductors and there are no "system complications", the interpretation problem reduces to one involving only FE emission physics and the electrostatics of the system geometry. We call systems of this kind *electronically ideal*. Systems where the interpretation problem cannot be reduced in this way (due to "system complications") are termed *electronically non-ideal*.

"System complications" can include (amongst other things): series resistance in the current path between the high-voltage generator and the emitter; voltage-deficit effects (due to emitter resistivity) that can lead to unexpected current-dependent variations in emitter characterization parameters; effects due to field emitted vacuum space charge; field-dependent changes in system geometry due to Maxwell stress; work-function changes due to current-related heating effects (Joule and/or Nottingham heating) and related desorption of adsorbates; and various effects related to field penetration into semiconductor emitters. With large area field electron emitters (LAFEs), which can involve large numbers of individual emission sites, plot non-linearity can occur because the emission comes from a distribution of emitters, all with different field enhancement factors. Further, several different complications can operate simultaneously.

This distinction between electronically ideal and non-ideal emitters can be set up in another way. The relationship between $F_C$ and the measured voltage $V_m$ can be written formally as

$$F_C = \frac{V_m}{\zeta_C}, \quad (1)$$

where $\zeta_C$ is a parameter called the *(characteristic) voltage conversion length (VCL)*. The VCL is a system characterization parameter, not a physical length. If the emitter physical condition is unchanging (apart from changes in applied field) and the VCL is constant, then the FE system is electronically ideal; if the VCL-value depends on current and/or voltage, then the FE system is electronically non-ideal.

It is possible for the system behaviour to be electronically ideal over part of the voltage range (usually the low-voltage part), but to be non-ideal over the remainder of the range.

If the FE system is electronically ideal, even if only over a limited range of input values (but a range of "reasonable length"), then values of emitter characterization parameters can be validly extracted from measured current–voltage data.

Two forms of *validity check* exist that can be applied in order to establish whether observed data have been taken from an electronically ideal system. First, a data-analysis plot of one of the three standard types (the Fowler–Nordheim plot is the best known) has to be "nearly linear" over all or part of its voltage range. If this test is passed, then good practice is to apply the so-called Orthodoxy Test [12]. If this second validity check is passed, then emitter characterization parameters can be validly extracted. If the Orthodoxy Test is failed, then characterization-parameter values extracted using standard data-analysis procedures are very likely to be spurious.

Until recently, the orthodoxy test was applied by using a spreadsheet, downloadable from the Royal Society (of London) website [12]. A couple of years ago, the lead author (MMA) felt that it would be more convenient for users if he built a webtool that would apply the Orthodoxy Test, and – if the test were passed – would extract relevant characterization parameters from the data [13,14].

In order to extract emitter characterization parameters from electronically ideal data-analysis plots, it is mathematically possible either to use (a simplified version of) 1928/29 FN FE theory or to use 1956 MG FE theory. The difference lies in the formulae that are used to interpret the extracted values of the slope and intercept of the straight line fitted to the plot (see [7] for an illustration). For slope-related parameters the difference in final values is typically 5% or slightly less, and not usually of any significance. However, for the area-like parameter extracted by using both the fitted slope and the fitted intercept, the two approaches yield final values typically varying by a factor of order 100. Since it is decisively known (see [8]) that the 1956 MG FE equation is "better physics" than the 1928/29 FN FE equation, it follows that (of the two choices) theory based on the MG FE equation should *always* be used for the data interpretation process, irrespective of which form of data analysis plot is being used. The theory used here and used in building the webtool is based on 1956 MG FE theory, and we do not allow the option of using the "worse physics" inherent in 1928/29 FN FE theory.

In this matter we have adopted a stance different from that found in much of existing FE technological literature, which uses a simplified version of 1928/29 FN FE theory to interpret the slope of a FN plot. Little direct harm is done by this, because usually interest is only in the determination of a dimensionless field enhancement factor from an extracted value of FN-plot slope, and the procedure error involved is usually less than 5%. But, in our view, scientific damage is done by the frequently repeated publication of an equation long known to be outdated (or, worse, the publication of a defective variant of it with a new type of error), and scientific damage is also done by the frequently repeated citation (often the only theoretical citation used) of a paper in which the discrepancy between theory and later experiment [15] can be shown to be very roughly of order $10^{15}$. The associated repeated failure to mention this discrepancy when citing the 1928 FN paper may tend to leave non-experts with the false view that "Fowler and Nordheim got the details right" in their 1928 paper, when in fact they did not. (By "repeated" we mean a failure that occurs in probably more than 1000 papers.) There is also, of course, an issue of repeated failure of the scientific Peer Review process.

Our view is that, in fact, there is no useful merit in using discredited equations that are nearly 100 years old in order to analyse modern FE data. For electronically ideal systems, the more modern approach described below provides both useful *additional* information for technology developers, and better value for money for FE research funders. Further, more widespread use of the Orthodoxy Test might help to diminish the numbers of questionable results in the literature. Thus, we hope that this paper can contribute in a small way to improving the quality of FE literature.

The inputs to both the spreadsheet and webtool versions of the Orthodoxy Test are an assumed value for the relevant local work function and the coordinates of two points (on a relevant data-analysis plot) that define the ends of a straight line fitted to the data. These two points are called the *(working) range limits*. There are four different pairs of quantities that can be used in a FE data-analysis plot, and three different conventional types of data-analysis plot. The decision was taken to make the webtool accommodate all twelve possible forms of data input, even though our strong view is: (a) that the best data-input form is the raw data that gives measured current as a function of measured voltage; and (b) that (at present) the best form of data-analysis plot is the Murphy–Good plot, discussed below.

The webtool is also set up so that it can accept and process data relating to *single-tip field electron emitters* (STFEs) (which are usually needle-like), and also data relating to *large area field electron emitters* (LAFEs). The term "LAFE" relates to any multi-emission-site device that has a significant macroscopic area (or "foot-print")—-often 1 to 25 $mm^2$, but not limited to this range. The characterization parameters most suitable for these two types of emitter are slightly different.

A main aim of this paper is to validate the software engineering of the webtool, but the paper also provides a further introduction to the tool and to the theory and reasons behind its development. This validation is best done using simulations of FE current as a function of voltage, because the exact form of the input is then known, and extracted outputs can be compared with the known inputs.

The structure of the remainder of the paper is as follows. Section 2 presents a thorough summary review of relevant basic emission theory (especially modern developments) and of the electronic engineering of FE current–voltage data interpretation. Section 3 presents a brief overview of the MMA webtool. Section 4 presents the results of simulations designed to test both the webtool and different methods of data analysis. Section 5 discusses conclusions and possible future developments.

Values of universal physical constants used in FE are given here to 7 significant figures, in field emission customary units. These, like SI units, are based on the modern system of equations that has $\epsilon_0$ in Coulomb's Law and that has (since 2009) been known as the *International System of Quantities*. These customary units are more convenient than SI units for discussing atomic-scale processes, and are recognized for continued use alongside SI units. Values of universal constants should be appropriately rounded in practical applications.

## 2. Review of basic theory

The theory here is based on the "scaled" form of Extended Murphy–Good FE theory, as developed by one of us (RGF). This is the most recent form of planar FE theory. We consider basic emission theory, the different forms of data input and data-analysis plot, the theory of the orthodoxy test, the different types of emitter characterization parameters that can be extracted from electronically ideal data-analysis plots, and the related extraction formulae.

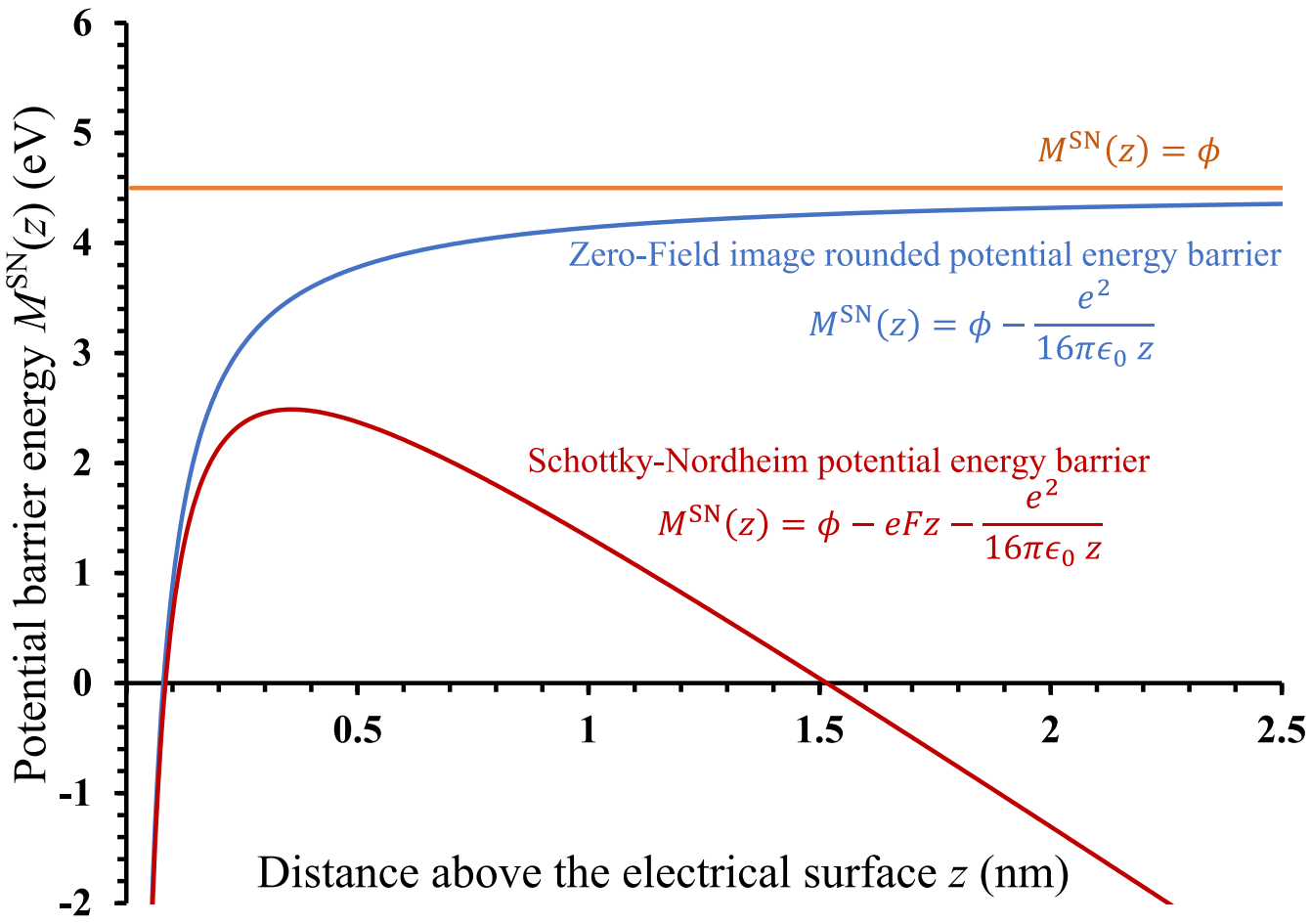

**Fig. 1.** Schematic diagram to show the image-rounded zero-field barrier and the Schottky–Nordheim (SN) barrier. The Schottky–Nordheim barrier is evaluated for $\phi$ = 4.50 eV and at $F \approx 2.8$ V/nm.

### 2.1. Summary review of Extended Murphy–Good emission theory

#### 2.1.1. The Schottky–Nordheim transmission barrier

In deriving the emission physics of CFE, two high-level factors need to be considered: the supply of electron current to the inside of the emitting surface, and the tunnelling probability (or, more generally, *transmission probability*) that an approaching electron will be emitted. The first of these factors is a long-solved standard problem in statistical mechanics (e.g., [16], see §11.31 to §11.36), so our remarks here concentrate on simple models for evaluating transmission probability, and aim to provide a simple "experimentalist-oriented" explanation of the physics involved. For simplicity, we now drop the subscript "L" from $\phi_L$, leaving it to be understood that $\phi$ denotes the local work function relevant to the emission location.

As already noted, CFE occurs when an intense negative electrostatic field, of magnitude often in the approximate range 3 to 5 V/nm (for a $\phi \approx 4.5$ eV emitting surface), is applied to the surface. As illustrated in Fig. 1, this changes the shape of the potential energy (PE) barrier from its zero-field "image rounded" shape to a reduced image-rounded PE barrier (e.g., [8], see Fig. 1), now often called the *Schottky–Nordheim (SN) barrier*.

Strictly, the form of the barrier is determined, not by the PE $U(z)$ alone (where $z$ is measured from the emitter's *electrical surface*), but by the difference $M(z) \equiv U(x) - E_n$, where $E_n$ is the electron's so-called *normal-energy*. $E_n$ is the component (of the electron's total-energy $E$) that is associated with electron motion normal to the emitter surface. Both $U(z)$ and $E_n$ need to be measured relative to the same energy reference level.

Strictly, also, the parameter $M(z)$ is not a potential energy, because it is a difference between a potential energy and a normal-energy. Following Newton's use of the term "motive force" in his original formulation of his Second Law [17] (see p.19 in the translated version [18] of *Principia*), $M(z)$ has sometimes been called the *electron motive energy*.

For a SN barrier of zero-field height equal to the local work function $\phi$, the motive energy $M^{SN}(z)$ for an electron with normal-energy equal to the Fermi level is given by:

$$M^{SN}(z) = \phi - eF_L z - \frac{e^2}{16\pi\epsilon_0 z}, \tag{2}$$

where $e$ is the elementary positive charge, and $\epsilon_0$ is the electrical permittivity of free space.

When the SN barrier becomes sufficiently narrow, cold electrons can quantum tunnel through to vacuum without requiring any additional energy. In this case, emission of electrons is in the CFE regime, and the electrostatic field must be intense enough (of magnitude at least around 3 V/nm for a $\phi$ = 4.50 eV surface) to make the SN barrier narrow enough for the cold electrons to be able to tunnel through. However, in practice, the field magnitude must stay well below a value where the emitter becomes unstable (due to heating or other effects) and self-destructs.

In the simplest form of tunnelling theory, the transmission probability $D^{SN}$ for the SN barrier is given by

$$D^{SN} \approx \exp\left[-(2\kappa_e)\int_{z_1}^{z_2}\{M^{SN}(z)\}^{1/2}\mathrm{d}z\right]. \tag{3}$$

where $(2\kappa_e)$ is an universal constant of value 10.24633 eV$^{-1/2}$nm$^{-1}$, and the limits of the integral are the points where $M^{SN}(z) = 0$. If we now denote the value of the integral in Eq. (3) by $Q^{SN}$, then this equation can be written in the simplified form

$$D^{SN} \approx \exp[-2\kappa_e Q^{SN}], \tag{4}$$

Fig. 2 shows how increasing the electrostatic field magnitude $F_L$: (a) affects the shape of SN barrier and decreases the SN barrier height; and (b) reduces the area $Q^{SN}$. Hence, from Eq. (4), increasing $F_L$ increases the transmission probability $D^{SN}$. At the reference field-magnitude (≈14.06 V/nm for a $\phi$ = 4.500 eV emitting surface) the barrier height and the integral $Q^{SN}$ both become zero.

For comparison, the behaviour of the exactly triangular (ET) barrier is also shown; the related mathematics follows the pattern of Eqs. (2) to (4), but with the image PE term removed.

The horizontal axis in Fig. 2 represents an electron normal-energy equal to the Fermi level. So, when the top of the barrier is pulled below the horizontal axis, as illustrated in Figs. 2(b) and 3(a), electrons with normal energy equal to the Fermi level (or above) can "fly over" the top of the barrier. This occurs at the *reference field-magnitude for the SN barrier* $F_R^{SN}$ given by

$$F_R^{SN} = (4\pi\epsilon_0/e^3)\phi^2 \equiv c_S{}^{-2}\phi^2 \approx (0.6944615\ \mathrm{V/nm})\cdot(\phi/\mathrm{eV})^2, \tag{5}$$

where $c_S$ [$\equiv \sqrt{e^3/(4\pi\epsilon_0)} \approx$ 1.999985 eV (V/nm)$^{-1/2}$] is the *Schottky constant*. For $\phi$ = 4.500 eV, $F_R^{SN} \approx$ 14.06 V/nm. This formula can be found, either by setting the maximum value of $M^{SN}(z)$ equal to zero, or by considering the case where $z_1 \rightarrow z_2$.

Using this definition, a parameter $f_C$ called the *characteristic scaled-field* (for a SN barrier of zero-field height $\phi$) is defined as the ratio

$$f_C \equiv \frac{F_C}{F_R^{SN}} = c_S{}^2\phi^{-2}F_C, \tag{6}$$

where $F_C$, as before, is a characteristic local field at some location "C" near the emitter apex (usually taken at the apex in modelling). The situation shown in Fig. 3(a) corresponds to $f_C$ = 1. As shown below, this parameter $f_C$ plays an important role in modern FE theory.

When electrons can easily fly over the top of a field-reduced barrier, the emission is no longer in the CFE regime. In fact, the derivation of the 1956 MG FE equation as given above breaks down at a $f_C$-value around 0.8. However, it is more useful to show a *regime diagram*: Fig. 3(b) shows the temperature/scaled-field regime where the derivation of a finite-temperature version of 1956 MG FE theory is adequately valid [19]. (The correction due to finite temperature is always small within this regime, typically less than 20%, and customary practice is to omit the temperature correction term from theory being used at room temperature, as has been done above.)

#### 2.1.2. The extended Murphy-Good equation for local emission current density

The 1956 zero-temperature Murphy–Good FE equation for characteristic local emission current density, $J_C^{MG0}$, can be written in the "linked" form

$$J_C^{MG0} = t_F^{-2} J_{kC}^{SN}, \tag{7}$$

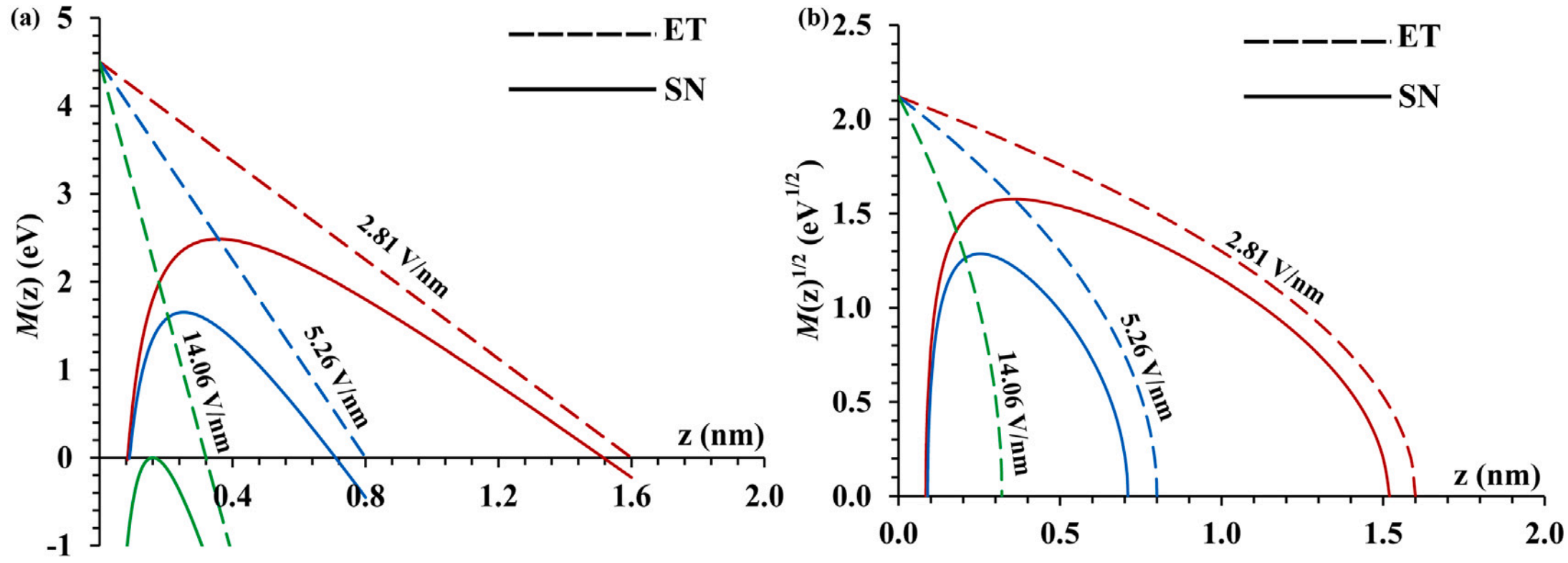

**Fig. 2.** To illustrate the changes, as the electrostatic field-magnitude $F_L$ increases, in: (a) the shapes of the exactly triangular (ET) and Schottky–Nordheim (SN) barriers, and (b) the areas $Q$ that appear in Eq. (4) for the SN barrier and in the equivalent equation for the ET barrier. The barriers are evaluated for $\phi = 4.500$ eV.

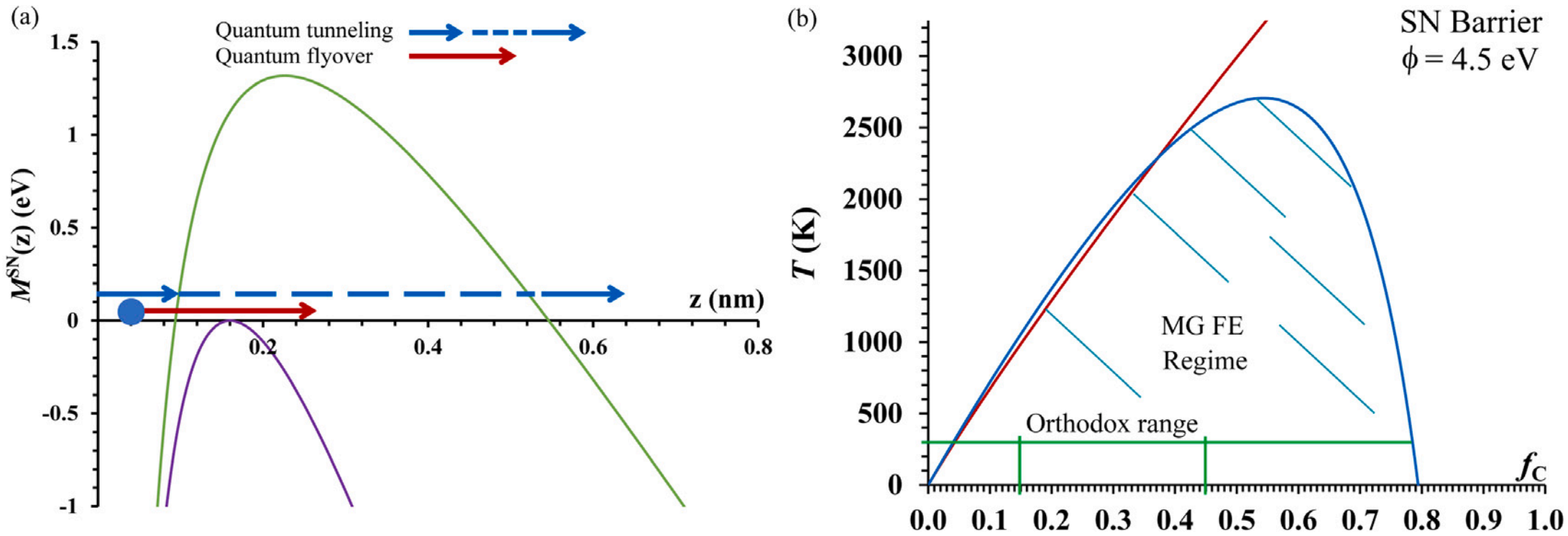

**Fig. 3.** (a) Schematic diagram to illustrate the distinction, in electron emission, between wave-mechanical (or"quantum") tunnelling and wave-mechanical (or"quantum") flyover. (b) *Regime diagram* that shows the regime of validity of the finite-temperature version of the 1956 Murphy–Good FE equation, as a function of temperature and characteristic scaled-field, for an electron-emitting surface of local work function 4.50 eV. The regime of validity is the region inside both the red (nearly straight) and the blue (hump-shaped) curves (which represent two different validity requirements). The green horizontal line shows the range of validity at 300 K.

$$J_{\mathrm{kC}}^{\mathrm{SN}} \equiv a\phi^{-1}F_{\mathrm{C}}^{2}\exp\left[-\frac{\mathrm{v_F}b\phi^{3/2}}{F_{\mathrm{C}}}\right], \tag{8}$$

where $a$ [ $\equiv e^3/(8\pi h_{\mathrm{P}}) \approx 1.541434 \times 10^{-6}$ A eV V$^{-2}$], and $b$ [ $\equiv 8\pi\sqrt{2m_{\mathrm{e}}}/(3eh_{\mathrm{P}}) \approx 6.830890eV^{-3/2}$ (V/nm)] are universal constants called the First and Second Fowler–Nordheim (FN) constants (see supplementary electronic material for Ref. [20]), $m_{\mathrm{e}}$ is the electron rest mass in free space, and $h_{\mathrm{P}}$ is Planck's constant. The parameters $\mathrm{v_F}$ and $\mathrm{t_F}$ are appropriate particular values (appropriate to a SN barrier defined by $\phi$ and $F_{\mathrm{C}}$) of well-known special mathematical functions (e.g., see [21]). The parameter $J_{\mathrm{kC}}^{\mathrm{SN}}$ is called the *kernel current density for the SN barrier* and is defined by eq. (8).

To define a suitable version of the so-called *Extended Murphy–Good (EMG) FE equation*, Forbes replaced Eq. (7) by the equation

$$J_{\mathrm{C}}^{\mathrm{EML}} = \lambda_{\mathrm{C}}J_{\mathrm{kC}}^{\mathrm{SN}}, \tag{9}$$

where $\lambda$ is a parameter originally called a "knowledge uncertainty factor" but now called a *prediction uncertainty factor*. $\lambda_{\mathrm{C}}$ is its value for location "C". The factor $\lambda$ was introduced into FE theory some years ago [22]; the name "EMG theory" is more recent [12].

As explained earlier, this parameter $\lambda_{\mathrm{C}}$ is a "placeholder" that *formally* takes account of ALL physical effects that are not included in 1956 MG FE theory, in particular the disregard of atomic-level effects. The 1956 MG FE theory pre-exponential $\mathrm{t_F^{-2}}$, and the temperature correction factor evaluated in the 1956 paper, are also swept into $\lambda$, because these are assumed to be very small corrections when compared with the major sources of uncertainty.

Neither the functional dependences of $\lambda$ nor its range of values is well known. Further, as research stands at present, neither FE theory nor FE experiment is good enough to derive reliable estimates of how it behaves: all one can do is to make informed guesses. Those of Forbes were based originally on the experience of Modinos, mainly via private discussions, with some tweaking (see [22]). The recent work of Lepetit [23] is also directly relevant (see [24]). The current suggestion by Forbes is that the prediction uncertainty factor is a function of field and most probably lies somewhere in the range $0.005 \leqslant \lambda_{\mathrm{C}} \leqslant 14$, but it would be no great surprise if the lower limit turned out to be pessimistically low.

### 2.1.3. Progress with the theory of the special mathematical function $\mathrm{v}(x)$

*Basic conventions*. In the period 2006–2010, significant advances in mathematical understanding were made in connection with the function "v" used in Murphy–Good FE theory. These are reviewed in [25]; the remarks here are a summary.

A major advance was to understand that "v" is in fact a very special solution of the Gauss Hypergeometric Differential Equation (HDE). This is the highest-level mathematical context in which "v" appears, and it follows that "v" should be expressed as a function of the independent variable in the Gauss HDE. A new convention has been to denote this variable by "$x$" and call it the *Gauss variable*.

An argument has been made that the mathematics of "v" should be conceptually separated from the use of "v" in specific modelling contexts such as field electron emission, and that $\mathrm{v}(x)$ should be treated as a special mathematical function (SMF) with its own body of mathematical theory. The symbols for SMFs are, by international convention, normally typeset upright, so we now prefer to write the function as $\mathrm{v}(x)$, and to also typeset upright the symbols for closely related functions, such as $\mathrm{t}(x)$. There seems no obvious objection-free short name for $\mathrm{v}(x)$, so it is provisionally being called the *principal field emission special mathematical function* (but "v" for short).

Previously, the mathematics of "v" has been formulated in terms of the *Nordheim parameter* $y$, which (it can be shown) is equal to $+\sqrt{x}$. It can be argued that, since $x$ is the natural variable for use in the mathematics, the legacy practice of using the Nordheim parameter $y$ should (from the strictly mathematical point of view) now be regarded as *mathematically perverse*. In particular, it is NOT normal mathematical practice to look for a solution to a differential equation in terms of a function of the SQUARE ROOT of the independent variable in the equation. Normal practice is to look for a solution that USES the independent variable in the differential equation.

It has also been argued that, in fact, the natural variable to use in an MG-type theory of current densities is the characteristic scaled-field $f_\mathrm{C}$ defined by Eq. (6), rather than the legacy convention of using the Nordheim parameter $y$. Both conventions should continue to be permissible, at least for the time being, but it can be argued that it is likely that experimentalists will in fact find that using $f_\mathrm{C}$ is a more powerful and flexible approach, particularly when discussing current–voltage measurements and theory.

Thus, our strongly recommended "21st Century approach" is that the special mathematical function $\mathrm{v}(x)$ should be applied to MG theory, as for example set out in Eq. (8), by setting $\mathrm{v}_\mathrm{F} = \mathrm{v}(x = f_\mathrm{C})$. The mathematical proof that this is a correct procedure is lengthy and is currently spread over several papers, using a variety of notations. A short argument that this is correct is as follows. In the legacy approach it has been shown that "$y$" in the modelling is the same parameter as "$y$" in the basic mathematics. It follows that "$y^2$" in the modelling is the same as "$y^2$" in the basic mathematics. It follows that if we replace "$y^2$" in the mathematics by $x$ and "$y^2$" in the modelling by $f_\mathrm{C}$, then the substitution procedure described above is a procedure compatible with the legacy approach. There remains a need for a full proper proof to be published in a single tutorial-type paper.

The alternative approach is the legacy modelling convention in which we write $\mathrm{v}_\mathrm{F} = \mathrm{v}(x = y^2)$. Eq. (8) is written using the symbol $\mathrm{v}_\mathrm{F}$ in order to allow either convention to be used.

*Scaled planar FE equations*. A further consequence of introducing the parameter $f_\mathrm{C}$ is that this allows the development of useful so-called *scaled equations* for kernel current densities. For the SN barrier, work-function-dependent *scaling parameters* for the exponent and pre-exponential, respectively, can be defined (using FE universal constants defined earlier) by

$$\eta(\phi) \equiv bc_\mathrm{S}^2 \phi^{-1/2}, \tag{10}$$

$$\theta(\phi) \equiv ac_\mathrm{S}^{-4} \phi^3. \tag{11}$$

Algebraic manipulation of Eq. (8), using these equations and also Eq. (6), yields the *scaled-format equation* for the kernel current density for the SN barrier, namely

$$J_\mathrm{kC}^\mathrm{SN} \equiv \theta f_\mathrm{C}^2 \exp\left[-\mathrm{v}(f_\mathrm{C}) \cdot \frac{\eta}{f_\mathrm{C}}\right], \tag{12}$$

Here, and below, the dependence of $\eta$ and $\theta$ on work-function is not normally shown explicitly, but the dependence of "v" on $f_\mathrm{C}$ is now shown explicitly.

A merit of this equation is that it contains only a single, direct, independent variable. This makes mathematical manipulations, including differentiation, markedly easier.

If this scaled equation is to be used to help interpret FE current–voltage measurements from electronically ideal systems, then a formula is needed that relates $f_\mathrm{C}$ to the measured voltage $V_\mathrm{m}$. This is achieved by using Eq. (21) below to define a parameter $V_\mathrm{mR}$, called the *reference measured voltage* (for the SN barrier), by

$$V_\mathrm{mR} = F_\mathrm{R}^\mathrm{SN} \zeta_\mathrm{C}, \tag{13}$$

where $\zeta_\mathrm{C}$ is (for an electronically ideal system) a system-specific characterization *constant* called the characteristic *voltage conversion length (VCL)*. For an electronically ideal system modelled using a SN barrier, $V_\mathrm{mR}$ is the measured voltage needed to pull the top of the SN barrier down to the Fermi level.

Applying a similar equation to the field-magnitude $F_\mathrm{C}$ yields

$$f_\mathrm{C} = \frac{F_\mathrm{C}}{F_\mathrm{R}^\mathrm{SN}} = \frac{V_\mathrm{m}/\zeta_\mathrm{C}}{V_\mathrm{mR}/\zeta_\mathrm{C}} = \frac{V_\mathrm{m}}{V_\mathrm{mR}}. \tag{14}$$

Thus, for an electronically ideal system, $f_\mathrm{C}$ is also "scaled measured voltage" (and, for a LAFE, is also "scaled macroscopic field").

*The "simple good approximation" for $\mathrm{v}(f_\mathrm{C})$*. As part of "21st Century" mathematical developments, several accurate (exactly equivalent) expressions, and some high-quality mathematical approximations, have been developed for $\mathrm{v}(x)$. These are described elsewhere [25]. Of relevance here is the so-called *simple good approximation* $\mathrm{v}_\mathrm{F06}$ [25,26]:

$$\mathrm{v}(f_\mathrm{C}) \approx \mathrm{v}_\mathrm{F06} = 1 - f_\mathrm{C} + \frac{1}{6} f_\mathrm{C} \ln(f_\mathrm{C}), \tag{15}$$

Over the range $0 \leqslant f_\mathrm{C} \leqslant 1$, where "v" takes values in the range $1 \geq \mathrm{v} \geq 0$, the maximum relative error in expression (15) is 0.33% and the maximum absolute error is 0.0024.

If this expression is inserted into Eq. (12), algebraic re-arrangement leads to the *expanded scaled format* for the SN-barrier kernel current density, namely

$$J_\mathrm{kC}^\mathrm{SN} \approx \theta f_\mathrm{C}^{(2-\eta/6)} \exp[\eta] \exp\left[-\frac{\eta}{f_\mathrm{C}}\right]. \tag{16}$$

As shown below, this equation forms the basis for the construction of Murphy–Good plots.

Note that the exponent $\exp[-\eta/f_\mathrm{C}]$ also appears in the elementary version of the 1928/29 FN FE equation. Thus, in this expanded scaled formulation, the whole of the difference between 1956 MG FE theory and elementary FE theory appears in the *pre-exponential* of the equations. This in turn affects the *intercept* of a data-analysis plot, and implies a need for accurate extraction of plot intercept values.

### *2.2. Data analysis plots and related issues*

*Data input variables.* As indicated above, our strong view is that by far the best choice for data input variables is to use the measured current and voltage and current $\{I_\mathrm{m}, V_\mathrm{m}\}$. This is because, for both ideal and non-ideal FE systems, these data are *experimental facts*, and are therefore scientifically valid items of information. However, other plot-variables are found in FE literature.

For a LAFE, the *macroscopic (or "LAFE-average") current density* $J_\mathrm{M}$ is defined by

$$J_\mathrm{M} \equiv \frac{I_\mathrm{m}}{A_\mathrm{M}}, \tag{17}$$

where $A_\mathrm{M}$ is the *macroscopic or "footprint" area* of the LAFE. This area $A_\mathrm{M}$ can be independently measured, so the macroscopic current density $J_\mathrm{M}$ is a well-defined experimental parameter.

Note that it is important that the subscript "M" (or "av") be added to the symbol for macroscopic current density. This is because, in real situations, emission comes only from the tips of individual emitters, and this "effective tip emission area" is only a small fraction of the "site area" (i.e., the footprint associated with a single emitter). Thus, the parameter $J_\mathrm{M}$ is much smaller than the characteristic local emission current densities ($J_\mathrm{C}$) discussed earlier, perhaps sometimes by a factor as much as $10^9$. Formal ways of dealing with this situation are discussed below.

In FE literature, this distinction between local current densities and macroscopic current densities is often not made, and the same symbol $J$ (and the same name "current density") are used for both: for $J_\mathrm{M}$ in diagrams and for $J_\mathrm{C}$ in equations. This can lead to publication situations where there are large apparent discrepancies between experiments and theory-as-given-in-the-paper. These discrepancies are often not discussed, apparently because no-one (authors, reviewers or editors) has detected their existence.

A further parameter widely used in LAFE literature is an (apparent) macroscopic field $F_{\mathrm{M}}^{\mathrm{app}}$ defined by

$$F_{\mathrm{M}}^{\mathrm{app}} = \frac{V_{\mathrm{m}}}{d_{\mathrm{M}}}, \tag{18}$$

where $d_{\mathrm{M}}$ is an experimental parameter associated with the FE system geometry. There is more than one geometrical option for the choice of $d_{\mathrm{M}}$, so for illustration we consider the commonly used situation of so-called *PPP geometry*, where a LAFE has been fabricated on one of a pair of well-separated parallel planar plates, and put $d_{\mathrm{M}} = d_{\mathrm{sep}}$, where $d_{\mathrm{sep}}$ is the separation between the plates.

In this case, what the experimentalist is trying to deduce is the true macroscopic field $F_{\mathrm{M}}^{\mathrm{true}}$ between the plates, which is defined by

$$F_{\mathrm{M}}^{\mathrm{true}} = \frac{V_{\mathrm{p}}}{d_{\mathrm{sep}}}, \tag{19}$$

where $V_{\mathrm{p}}$ is the voltage *between the front surface of the emitter substrate and the distant counter-plate*. However, in real systems there is no requirement for $V_{\mathrm{P}}$ to be equal to the measured voltage $V_{\mathrm{m}}$. This is especially the case when a LAFE is fabricated on the surface of a resistive semiconductor slab or layer, because it is the *front* surface of the slab or layer that has to be treated as one of the parallel plates.

Thus, the experimentalist thinks that he/she is calculating $F_{\mathrm{M}}^{\mathrm{true}}$ but is actually calculating $F_{\mathrm{M}}^{\mathrm{app}}$. The correct formula for $F_{\mathrm{M}}^{\mathrm{true}}$ has the form

$$F_{\mathrm{M}}^{\mathrm{true}} = \frac{V_{\mathrm{m}}}{Z_{\mathrm{mM}}}, \tag{20}$$

where the so-called *measured-voltage-to-macroscopic-field conversion parameter* $Z_{\mathrm{mM}}$ is defined by this equation, and is a system characterization parameter, rather than a physical distance. For an electronically ideal FE system, $Z_{\mathrm{mM}}$ is indeed a constant equal to a physical distance in the system; but for an electronically non-ideal system $Z_{\mathrm{mM}}$ is an initially unknown characterization parameter that is likely to be a function of voltage and/or of current. For a more mathematical discussion, see [27].

A consequence of this situation is that experimental voltage data that are converted using Eq. (18) lose their validity, and data plots based on this data cease to be “experimental data” but become “unconfirmed pre-converted mathematical data”. After validity checks have been applied, these unconfirmed data can either be “recognized” as true experimental data (if the checks are passed), or may remain in a state of ambiguity (if the checks are failed). This is because some types of system complication (e.g., high series resistance) will certainly generate false experimental data plots, but other types of complication (e.g., space-charge) may be generating true experimental data for an electronically non-ideal FE system. In the present state of research knowledge, it can be difficult to distinguish between these alternative possibilities.

In FE literature these difficulties are usually not discussed, and often all data plots are presented as if they were true experimental data. Non-expert readers, in particular industrialists and defence scientists interested in using field emitter characterization data, may not realize that authors, reviewers and editors may (sometimes or often) not be able to distinguish between false experimental FE data and true experimental FE data. A merit of the Orthodoxy Test, and of our webtool implementation of it, is that they can (with good probability) identify which data plots in the literature are true experimental data taken from electronically ideal FE systems.

Another unfortunate feature of FE experimental/technological literature is that sometimes pre-conversion of $I_{\mathrm{m}}(V_{\mathrm{m}})$ data into $J_{\mathrm{M}}(F_{\mathrm{M}}^{\mathrm{app}})$ data is done, but the papers concerned do not record the values of both of the parameters ($A_{\mathrm{M}}$ and $d_{\mathrm{M}}$) used to make the conversion. If the resulting data plot fails the orthodoxy test, then the original raw experimental data cannot easily be retrieved. The technology may be interesting but the emitter materials cannot be characterized by readers unless they contact the author(s).

The ambiguities just described, over the meanings of the terms “current density” and “macroscopic field”, are a primary cause of our strong view that by far the best data-plotting approach is $I_{\mathrm{m}}(V_{\mathrm{m}})$. In this case the ambiguities do not arise. And, as already stated, whether or not the FE system turns out to be electronically ideal, at least the plotted data are true raw experimental data.

These ambiguities have also been a partial cause of the decision to build the webtool so that it can accept all four variants of data-input form, namely $I_{\mathrm{m}}(V_{\mathrm{m}})$, $I_{\mathrm{m}}(F_{\mathrm{M}}^{\mathrm{app}})$, $J_{\mathrm{M}}(V_{\mathrm{m}})$ and $J_{\mathrm{M}}(F_{\mathrm{M}}^{\mathrm{app}})$. In practice, needle-type emitters are nearly always analysed using the $I_{\mathrm{m}}(V_{\mathrm{m}})$ approach. With LAFEs, all four of the approaches have sometimes been used, but the $J_{\mathrm{M}}(F_{\mathrm{M}}^{\mathrm{app}})$ approach is probably the most common.

*The conventional forms of data-analysis plot*. In order to keep discussion general, we shall here use “$X$” to denote the independent variable in the data input (either $V_{\mathrm{m}}$ or $F_{\mathrm{M}}^{\mathrm{app}}$), and “$Y$” to denote the dependent variable ($I_{\mathrm{m}}$ or $J_{\mathrm{M}}$).

All three of the conventional forms of FE data-analysis plot have the mathematical form of a plot of $\ln\{Y/X^{n}\}$ versus $1/X$, where $n$ is a parameter called here the *data plot index*, and the “curly-bracket notation" $\{Z\}$ means “take the numerical value of $Z$ when $Z$ is expressed in the designated units, discussed in the related text”. There is no requirement for $n$ to be integral.

In some cases common logarithms (to base 10) have been used by experimentalists for making plots and evaluating plot slopes. The webtool will convert common logarithms to natural logarithms (to base e). The theory below is based on the use of natural logarithms. It is also highly desirable that any new data-analysis plots are made using only the SI units A, V and m (not sub-multiples of these units). The Orthodoxy Test will work if any consistent set of units has been used, but evaluation of characterization parameters requires the consistent use of only A, V and m (in particular, fields must be in V/m and current densities in A/m$^2$). If necessary, conversion of units must be undertaken before data entry into the webtool.

The Millikan–Lauritsen (ML) plot [28] has $n = 0$, and is based on the empirical law found experimentally by Lauritsen in his PhD work [29] and reported in [30]. The Fowler–Nordheim (FN) plot has $n = 2$, and was introduced by Stern et al. [11] because the 1928/29 FN FE theory of the $J_{\mathrm{L}}(F_{\mathrm{L}})$ dependence predicts that this form will generate a straight-line plot. The Murphy–Good (MG) plot has $n = \kappa^{\mathrm{SN}} = 2 - \eta/6$, and was introduced by Forbes [7] because the expanded scaled version of the MG theory of the $J_{\mathrm{L}}(F_{\mathrm{L}})$ dependence predicts that this form will generate a “very nearly straight” line.

Fig. 4(a) uses EMG theory to present a simulated current–voltage characteristic $I_{\mathrm{m}}(V_{\mathrm{m}})$ for the “typical” metal local work-function value $\phi = 4.50$ eV, and for illustrative characterization parameters $\zeta_{\mathrm{C}} = 200$ nm and $A_{\mathrm{fC}}^{\mathrm{SN}} = 100$ nm$^2$ (These values have been chosen as typical extracted values for a tungsten STFE). Fig. 4(b) presents simulations of the three conventional types of data-analysis plot, based on the data shown in Fig. 4(a).

Other data-analysis-plot forms have been suggested, based in part on the thinking that the mathematical form of the experimental dependence of $I_{\mathrm{m}}$ on $V_{\mathrm{m}}$ does not have to be the same as the mathematical form of the theoretical dependence of $J_{\mathrm{L}}$ on $F_{\mathrm{L}}$ in SPME methodology. Thus, for example, Abbott and Henderson [31] suggested $n = 3$ or $n = 4$, and Forbes, Popov et al. [32] have suggested that the actual experimental value of $n$ should be determined by “best fit” methods.

Notwithstanding this, the present webtool and paper are confined to analysis of the three conventional data-analysis plot forms, because these are the ones that have normally been used in experimental papers.

It should be added that eventually the subject area will need to move on to the use of data-analysis tools that take local surface curvature into account in emission theory, probably based in the first instance on the work of Kyritsakis and Xanthakis [33,34]. But it has appeared to us that a useful (and probably necessary) prior task is to develop improved practice within the framework of SPME methodology.

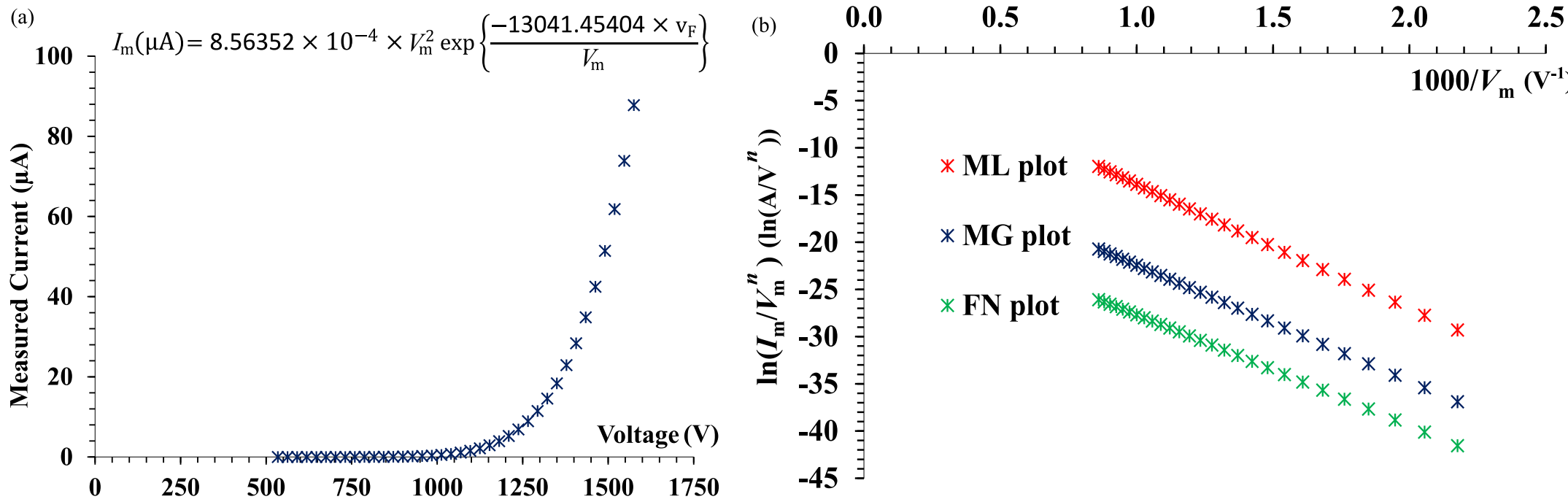

**Fig. 4.** Simulated plots for: (a) current–voltage characteristics for $\phi = 4.50$ eV, $\zeta_C = 200$ nm, $A_{fC}^{SN} = 100$ nm$^2$; and (b) the related data analysis plots, with $n = 0$ for ML plots, $n = 2$ for FN plots and $n = 1.22719$ for MG plots.

### 2.3. Emitter characterization parameters

We first need to make the point that there are two categories of emitter characterization parameter: those obtained by carrying out data-plot analysis after validity checks have been passed; and those obtained by applying standard "ideal-emitter" procedures to emitters that have not been tested or have failed the orthodoxy test or equivalent validity check. In the former case, provided that there are no distorting effects due to counter-electrode adjacency, the extracted parameters characterize the emitter *material*, as fabricated in a specified way. In the latter case the extracted parameters do NOT reliably characterize the emitter or the emitter material, but might provide a useful laboratory record of how a particular emitter, prepared and mounted in a particular way, in a particular FE system, was behaving on a particular day. The remarks here apply to emitters that have passed validity checks.

*Slope-related parameters*. Parameters extracted from the data-plot slope alone relate to the geometry of the system, as it affects the system electrostatics. Our view is that scientifically the most satisfactory approach is to suppose that there exists a formula $I_m(F_C)$ for predicting the measured current $I_m$ as a function of a well-specified characteristic LOCAL field $F_C$, and then write a formula that connects $F_C$ to the measured voltage $V_m$. As indicated earlier, we prefer the form

$$F_C = \frac{V_m}{\zeta_C}, \tag{21}$$

where $\zeta_C$ is called the *characteristic voltage conversion length (VCL)*. In general, $\zeta_C$ is a system characterization parameter that is a function of measured voltage and/or measured current, but for electronically ideal systems $\zeta_C$ becomes a constant. For ideal systems, a systematic review of the related quantitative electrostatics is in preparation (de Assis, Dall'Agnol and Forbes, in preparation, 2022).

In FE literature, the reciprocal of $\zeta_C$ is denoted by $\beta$ and is sometimes used instead of the VCL. We avoid this approach, partly because of potential confusion with the widespread FE convention that uses $\beta$ to denote the field enhancement factor discussed below, partly because we think it easier to use a parameter typically measured in "nm", rather than one typically measured in "$m^{-1}$".

In FE literature relating to LAFEs, it is common practice to attempt to describe the electrostatics of FE systems by using one of a set of related dimensionless parameters called here characteristic *field enhancement factors (FEFs)*. Parameters of this kind are denoted here by $\gamma_{MC}$ and defined by

$$\gamma_{MC} \equiv \frac{F_C}{F_M^{true}}. \tag{22}$$

We use the symbol "$\gamma$", rather than the more conventional symbol "$\beta$", in order to avoid confusion with the use of $\beta$ to denote the reciprocal of a VCL.

If the experimental $I_m(V_m)$ data plot passes validity checks (including the Orthodoxy Test), then formulae discussed below (depending on the chosen data-plot form) can be used obtain an extracted value $\zeta_C^{extr}$ of the characteristic VCL.

For LAFEs, an extracted value of the related FEF $\gamma_{MC}$ can then be obtained from the formula

$$\gamma_{MC}^{extr} = \frac{d_M}{\zeta_C^{extr}}, \tag{23}$$

where $d_M$ is the parameter used earlier and called the *macroscopic distance*.

There are, in fact, different geometrical options for the choice of macroscopic distance, and there exist correspondingly different types of dimensionless FEF. The main alternatives currently used are PPP geometry (where the macroscopic distance is the plate separation, as discussed earlier), and "gap geometry", where the macroscopic distance is taken equal to the gap length between the emitter apex and a closely adjacent counter-electrode (which is sometimes a pointed probe). For clarity, the definition of a dimensionless PPP-geometry FEF is illustrated in the Appendix.

Care also needs to be taken over how macroscopic distance is defined when considering, for example, cylindrical-wire geometry or blade geometry (likely to be relevant for emission from graphite flake edges [35], for example). The formulae given below should work adequately for all types of macroscopic distance, but what type of FEF you get out depends on what type of macroscopic distance you put in. Note that comparing the numerical values of different *types* of FEF (which is sometimes done in FE literature) is not electrostatically legitimate, except in a general qualitative fashion.

When working with scaled equations, the characterization parameters "reference measured voltage" and "reference macroscopic field" can be obtained in a generally similar manner, but extraction of these parameters is not currently implemented in the webtool.

Finally, we note that in FE literature a substitution of the form

$$F_C = \frac{V_m \gamma_{MC}}{d_M}, \tag{24}$$

is sometimes made into an equation for $I_m(F_C)$, and it is implicitly assumed that $d_M$ is a physical distance, that $F_M$ is a true macroscopic field, and that $\gamma_{MC}$ is a well-defined constant. These things are true for an electronically ideal system, but they cannot all be true for a non-ideal system. There is an awkward (unsolved) pedagogical problem of how best to formulate the theory in such cases. At present it looks as though it may be best to have parameters that represent the "ideal" values, and a set of correction factors, as for example done in [36]. Further discussion of this topic is outside the scope of this paper.

*Area-like parameters*. The definition and measurement of area-like parameters via FE experiments is very complicated. The approach here is a slight advance on previous treatments, but should be regarded as "work in progress".

In order to make a prediction about the emission current from a needle-like or post-like emitter, it is first necessary to adopt some specific model for the emitter shape and for how the local work-function varies across the model surface. As before, we assume here that the work function is uniform across the surface.

Ideally, one would wish to have a good theoretical expression for the local emission current density (LECD) and integrate this across the emitter surface. But this cannot be done at present, because of the unknown nature of the prediction uncertainty factor in Eq. (9). Hence, instead, one makes the planar emission approximation and integrates the kernel current density for the SN barrier over the surface of the model. This yields a *notional emission current* (for the SN barrier) $I_{\mathrm{n}}^{\mathrm{SN}}$ that one can write in the form

$$I_{\mathrm{n}}^{\mathrm{SN}} \equiv A_{\mathrm{nC}}^{\mathrm{SN}} J_{\mathrm{kC}}^{\mathrm{SN}}, \tag{25}$$

where, as before, $J_{\mathrm{kC}}^{\mathrm{SN}}$ is the value of the kernel current density at some suitably chosen characteristic location. The parameter $A_{\mathrm{nC}}^{\mathrm{SN}}$ is defined by this equation and is called the *notional emission area* (for the assumed emitter shape, etc., model, as analysed using a SN barrier). The value of this notional area will depend on the details of the assumed emitter model, and the choice of location "C".

We now assume that (for a given emitter model) if, we knew an exact expression for the LECD, then that would yield a "true model emission current" $I_{\mathrm{tm}}^{\mathrm{SN}}$ that could be written in the form

$$I_{\mathrm{tm}}^{\mathrm{SN}} = \lambda_J I_{\mathrm{n}}^{\mathrm{SN}}, \tag{26}$$

where $\lambda_J$ is a prediction uncertainty factor (of the same general kind as considered before) that is associated with uncertainties in emission theory. It will depend on many things, and we do not think it helpful to introduce complicated notation about this. The symbol is basically a "placeholder" that records the fact that uncertainty exists.

Unfortunately, this is not the end of the story because, when dealing with a real emitter, it may be that the real emitter and its surface condition do not correspond to the assumptions made in the model. This introduces a second source of uncertainty, and a formula for the "predicted measured current" $I_{\mathrm{p}}^{\mathrm{SN}}$ has to be written

$$I_{\mathrm{p}}^{\mathrm{SN}} = \lambda_{\mathrm{EM}} I_{\mathrm{tm}}^{\mathrm{SN}} = \lambda_{\mathrm{EM}} \lambda_J I_{\mathrm{n}}^{\mathrm{SN}} = \lambda_{\mathrm{EM}} \lambda_J A_{\mathrm{nC}}^{\mathrm{SN}} J_{\mathrm{kC}}^{\mathrm{SN}}, \tag{27}$$

where $\lambda_{\mathrm{EM}}$ is a second prediction uncertainty factor associated with deficiencies in the emitter model.

This formula can be simplified by defining a new area-like parameter $A_{\mathrm{fC}}^{\mathrm{SN}}$, called the *formal emission area*, by

$$A_{\mathrm{fC}}^{\mathrm{SN}} = \lambda_{\mathrm{EM}} \lambda_J A_{\mathrm{nC}}^{\mathrm{SN}}. \tag{28}$$

Eq. (27) can then be rewritten in the simplified form

$$I_{\mathrm{p}}^{\mathrm{SN}} = A_{\mathrm{fC}}^{\mathrm{SN}} J_{\mathrm{kC}}^{\mathrm{SN}}. \tag{29}$$

We now assert that a formula for actual measured current $I_{\mathrm{m}}$ can also be written in this form, as

$$I_{\mathrm{m}} = A_{\mathrm{fC}}^{\mathrm{SN}} J_{\mathrm{kC}}^{\mathrm{SN}}. \tag{30}$$

Since $I_{\mathrm{m}}$ is well-defined, and the expression for $J_{\mathrm{kC}}^{\mathrm{SN}}$ is well defined (though there will be some uncertainty over the values of $\phi$ and $F_{\mathrm{C}}$), it follows that in principle $A_{\mathrm{fC}}^{\mathrm{SN}}$ is a well-defined parameter and that one might hope to extract its value from experiment.

From Eq. (17) earlier, it follows that (for LAFEs)

$$J_{\mathrm{M}} = \frac{I_{\mathrm{m}}}{A_{\mathrm{M}}} \approx J_{\mathrm{kC}}^{\mathrm{SN}} \left( \frac{A_{\mathrm{fC}}^{\mathrm{SN}}}{A_{\mathrm{M}}} \right) \equiv \alpha_{\mathrm{fC}}^{\mathrm{SN}} J_{\mathrm{kC}}^{\mathrm{SN}}, \tag{31}$$

where $\alpha_{\mathrm{fC}}^{\mathrm{SN}}$ [$\equiv A_{\mathrm{fC}}^{\mathrm{SN}}/A_{\mathrm{M}}$] is the *formal area efficiency* (for the SN barrier) and is defined by this equation. This dimensionless parameter is a measure of what fraction of the LAFE footprint area is actually emitting electrons. Values of $\alpha_{\mathrm{fC}}^{\mathrm{SN}}$ are not well known, but are thought to typically lie in the range $10^{-9}$ to $10^{-4}$.

The area-like parameters that are extracted from FE experiments are these *formal* parameters. *For a real field emitter there is no easy relationship between the extracted formal emission area and measures of the real emitting area.*

It is also possible to define a *notional area efficiency* (for the SN barrier) by the relation $\alpha_{\mathrm{nC}}^{\mathrm{SN}} = A_{\mathrm{nC}}^{\mathrm{SN}}/A_{\mathrm{M}}$. This parameter cannot be measured but can be estimated theoretically [37], and (when combined with the prediction uncertainty factor $\lambda_{\mathrm{C}}^{\mathrm{SN}}$) can perhaps give a rough indication of an upper limit on the likely range of values of $\alpha_{\mathrm{fC}}^{\mathrm{SN}}$.

*Commentary*. There is an argument to be made that field electron emission is still in a pre-scientific or "partially scientific" state. This is because FE does not really satisfy either of the requirements needed for a subject area to be regarded as "properly scientific", namely: (a) compatible with and deducible from general principles of physics; and/or justified by experiment. A subject area where virtually all theory disregards the existence of atoms fails on the first test. It also seems to be true that at no stage in the 100 years of the subject's existence have quantitative comparisons of theory and experiment been made that are precise, reliable and decisive (although, of course, general trends are compatible). Further, defining and precisely measuring/calibrating the relevant *real* surface electrostatic field is a major problem, not yet adequately solved.

A conclusion drawn in earlier work was that attempting to make direct comparisons between theory and experiment would not provide an easy way of putting FE onto a better scientific basis. (If this were so, then reliable comparisons would have been made by now.) Rather, it was concluded (Forbes, unpublished work) that the scientific endeavour related to making FE into "proper science" needs to be split into four parts: (1) activities aimed at reducing the massive amount of confusion and error in FE experimental and technological literature, and at establishing a "common 21st Century starting place" for developing future FE science (including activities aimed at persuading people to go there); (2) investigation and development of methods (including the design of suitable apparatus) for the accurate extraction of formal emission area and other relevant parameters from experiments; (3) theoretical investigation of what needs to be done in order to develop improved theory (some of the problems seem to be very deep); and (4) investigation of any other procedures that can help reduce (or "side-step") the prediction uncertainty issues just discussed. To these one should perhaps add: (5) establishing how to measure, reliably and precisely, the relevant local surface electrostatic field.

Obviously, this paper contributes to the second of these tasks, and also aims to help reduce errors in current FE literature.

### 2.4. Basic theory of the Orthodoxy Test

The FE Orthodoxy Test is a powerful validity check that can be applied to any of the three conventional forms of data-analysis plot, whatever the choice of data-input variables, and whether or not authors have remembered to label the axes on their data plots clearly. It should be seen as an "engineering triage test" that assesses whether emitter characterization parameters deduced using conventional data-analysis techniques are likely to be valid, and divides data plots into three broad classes: "Pass" - extracted results are likely to be reliable; "Fail" - extracted results are highly likely to be spurious; and "Inconclusive" - reliability is certainly not guaranteed and further investigation is needed. The test has a good scientific basis, but should not be treated as precise science. It should be understood as a test AGAINST the hypothesis that the data in question have been taken from an electronically ideal system that can be adequately modelled by MG FE theory, using the value of local work function that is input.

The test is based on the scaled form of Murphy–Good FE theory, and works by extracting from the experimental data the range of scaled-field values ($f_{\mathrm{C}}$-values) that would be deduced from the experimental data-input variable if the system were electronically ideal and MG theory applied. This extracted range is then compared for compatibility with *reference zones* for $f_C$, for the three output categories described above. These reference zones depend on the assumed local work-function of the emitter, and are determined in a manner described below.

| (a) | Zone definitions for Orthodoxy Test | |
|---|---|---|
| | Boundaries | $f_C$-range for $\phi$ = 4.50 eV |
| F: | $f_C < f_1$ | < 0.10 |
| I: | $f_1 \leq f_C < f_2$ | 0.10 \| 0.15 |
| P: | $f_2 \leq f_C \leq f_3$ | 0.15 \| 0.45 |
| I: | $f_3 < f_C \leq f_4$ | 0.45 \| 0.75 |
| F: | $f_4 < f_C$ | > 0.75 |

| (b) | Orthodoxy Test conditions |
|---|---|
| P: | If the whole extracted range is within the "PASS" (green) zone, the test is PASSED |
| F: | If any part of the extracted range is within either of the "FAIL" (red) zones then the test is FAILED |
| I: | In all other circumstances the test is INCONCLUSIVE, and further investigation is needed |

**Fig. 5.** Orthodoxy test summary: (a) definitions of zone boundaries; (b) definitions of "Pass", "Fail" and "Inconclusive" conditions.

For the FE system under test, the working range is determined by extracting individual $f_C$-values that correspond to the lower ($f_{low}$) and upper ($f_{up}$) ends of the working range, using the coordinates $\{\bar{X}, L\}$ of the ends of a straight line fitted to the experimental data. ($\bar{X}$ is an alternative symbol for $X^{-1}$; as before, "$X$" is being used here to denote the data-input variable.) The $\bar{X}$-values used must correspond exactly to the those of the lowest and highest data-points used. Also, bear in mind that $f_{low}$ is derived from the right-hand end of the plot, and $f_{up}$ from the left-hand end.

The slope $S^{fit}$ depends on which type of data plot is being used, but the extraction formula does not depend on which data-variables are being used.

For a Fowler–Nordheim (FN) plot, the extraction formula is [12]

$$f_C^{extr} = \frac{s_t \cdot \eta(\phi)}{|S_{FN}^{fit}| \cdot X^{-1}}, \tag{32}$$

where $S_{FN}^{fit}$ is the (negative) slope of the line fitted to the FN plot, and $s_t$ [$\equiv s(f_t)$] is the so-called "fitting value" of the slope correction function $s(f)$. It is usually adequate to approximate $s_t \approx 0.95$ [12].

For a Murphy–Good (MG) plot, the extraction formula is [7]

$$f_C^{extr} = \frac{\eta(\phi)}{|S_{MG}^{fit}| \cdot X^{-1}}, \tag{33}$$

where $S_{MG}^{fit}$ is the (negative) slope of the line fitted to MG plot. Application of the Orthodoxy Test to MG plots has been discussed in more detail in [38].

For a Millikan–Lauritsen (ML) plot the FN-plot formula is used, but with a (negative) slope value $S_{FN}^{eff}$ given by [14]

$$S_{FN}^{eff} \approx S_{ML}^{fit} + \frac{4}{\bar{X}_{left} + \bar{X}_{right}} \tag{34}$$

where $\bar{X}_{left}$ and $\bar{X}_{right}$ are the horizontal-axis coordinates of the left-hand and right-hand range limits of an experimental ML plot.

Fig. 5(a) below shows fives *zones* of extracted values that are relevant to the three Orthodoxy-Test outcomes (F, I, P). The *zone boundaries* are denoted (in ascending order) by $f_1 < f_2 < f_3 < f_4$. These boundaries depend on the assumed local work function, as shown in Table 1. For convenience, the values for $\phi = 4.50$ eV are also shown in Fig. 5. This useful "traffic light" approach to the Orthodoxy Test was devised by Dr Eugeni Popov and colleagues, of the Ioffe Institute in Saint-Petersburg.

### 2.5. Extracting emitter characterization parameters for electronically ideal FE systems

#### 2.5.1. Which parameters from which plot?

We next discuss the detailed methods by which characterization parameters are extracted from the conventional plot types, using the various forms of data input, when the FE system under investigation has passed validity checks. Each of the 12 possible combinations of data-input variables and data-plot form requires a slightly different approach, and the webtool accordingly has twelve sets of formulae available for use.

**Table 1**
The Orthodoxy Test zone boundaries as functions of the local work-function $\phi$.

| $\phi$ (eV) | $f_1$ | $f_2$ | $f_3$ | $f_4$ |
|---|---|---|---|---|
| 5.5 | 0.09 | 0.14 | 0.41 | 0.69 |
| 5.0 | 0.10 | 0.14 | 0.43 | 0.71 |
| 4.5 | 0.10 | 0.15 | 0.45 | 0.75 |
| 4.0 | 0.11 | 0.16 | 0.48 | 0.79 |
| 3.5 | 0.11 | 0.17 | 0.51 | 0.85 |
| 3.0 | 0.12 | 0.18 | 0.54 | 0.91 |
| 2.5 | 0.13 | 0.20 | 0.59 | 0.98 |

**Table 2**
The four data-input formats and their related extracted and derived parameters.

| Data input format | Extracted parameters | | Main derived parameters | |
|---|---|---|---|---|
| | From slope | From slope and intercept | Using $d_M$ | Using $A_M$ |
| $I_m(V_m)$ | $\zeta_C$ | $A_{fC}^{SN}$ | $\gamma_C$ | $\alpha_{fC}^{SN}$ |
| $I_m(F_M)$ | $\gamma_C$ | $A_{fC}^{SN}$ | $\zeta_C$ | $\alpha_{fC}^{SN}$ |
| $J_M(V_m)$ | $\zeta_C$ | $\alpha_{fC}^{SN}$ | $\gamma_C$ | $A_{fC}^{SN}$ |
| $J_M(F_M)$ | $\gamma_C$ | $\alpha_{fC}^{SN}$ | $\zeta_C$ | $A_{fC}^{SN}$ |

The whole process of FE system characterization would be much simpler (and, arguably, more accurate) if the community could agree on a single well-defined procedure that was not prone to error. As already indicated, our view is that (within SPME methodology) the obvious choice at present is the raw $I_m(V_m)$ experimental data analysed using as Murphy–Good plot. Thus, we shall only look at the fine details of how to process $I_m(V_m)$ data. (The other forms of data-input require broadly similar procedures.)

With needle-geometry emitters, as used in electron microscopes and traditional projection-type field electron microscopes, an $I_m(V_m)$ approach is the only sensible one, since there is significant geometrical difficulty in defining what might be meant by "true macroscopic field". In the $I_m(V_m)$ approach, using the "non-scaled" Eqs. (8) and (9), the parameter directly extracted from the plot slope is the voltage conversion length $\zeta_C$ (or its reciprocal, which we denote here by $\beta_V$). A related parameter is the so-called (characteristic) *shape factor* or "field factor" $k_C$ (also called a "$k$-factor"), defined via

$$\zeta_C = k_C r_a, \tag{35}$$

where $r_a$ is the emitter's apex radius of curvature.

This approach can also be applied to LAFEs, in which case the derived parameters apply to the sharpest emitters in the array. For LAFEs, a characteristic macroscopic field enhancement factor $\gamma_{MC}$ can be derived using Eq. (23); this also applies to the sharpest emitters in the array. For LAFEs, the question of how the extracted VCL and FEF values relate to the actual distribution of such parameters for the emitters in the array is exceptionally complicated in detail and is outside the scope of this paper.

If, instead of the "non-scaled equations", a scaled form of equation is used to interpret the plot, then the parameter directly extracted from the plot slope is the reference measured voltage $V_{\mathrm{mR}}$. This can be converted to the corresponding VCL (and thence to other forms of characterization parameter) via the relation

$$\zeta_{\mathrm{C}} = c_{\mathrm{S}}^2 \phi^{-2} V_{\mathrm{mR}}, \tag{36}$$

This equation is derived using Eqs. (5), (13) and (14).

With the dependent variable ($I_{\mathrm{m}}$ or $J_{\mathrm{M}}$), the situation is much simpler. If $I_{\mathrm{m}}$ is used then the extracted parameter is the (characteristic) formal emission area (for the SN barrier) $A_{\mathrm{fC}}^{\mathrm{SN}}$. If $J_{\mathrm{M}}$ is used then the extracted parameter is the (characteristic) formal area efficiency (for the SN barrier) $\alpha_{\mathrm{fC}}^{\mathrm{SN}}$. (But many FE technologists apparently do not understand this, due to the widespread use of a defective current-density equation in FE technological literature, as discussed earlier.) In both cases, the other characterization parameter is obtained via Eq. (31).

The information just discussed is presented, in a more precise form, on the first line of Table 2. Equivalent information for the other three choices of data-input parameter are shown on the remaining lines of the table.

### 2.5.2. Extraction formulae

In general, there are several different methodologies available for extracting characterization-parameter values from data-analysis plots. Our approach, when designing the webtool, was to use the so-called "tangent method" described below. For a more general discussion see [39] or (better) https://doi.org/10.13140/RG.2.2.32112.81927/3. This Section discusses some of the detailed formulae used in the webtool.

*Fowler–Nordheim plots: slope analysis*. For the extraction of parameters from the FN plot slope, the extraction theory has been written out in full, partly as an example, partly because it is the commonest analysis undertaken in the literature, but the proof of the slope formula is not well known and is not well presented in existing literature. In other cases only the final formulae will be given here.

With FN plots, it is convenient to work with non-scaled versions of MG FE theory. Combining Eqs. (8) and (29) yields the "direct" (i.e., "non-scaled") equation for measured current in Extended Murphy–Good theory, namely

$$I_{\mathrm{m}} = \{A_{\mathrm{fC}}^{\mathrm{SN}} a \phi^{-1} \zeta_{\mathrm{C}}^{-2}\} V_{\mathrm{m}}^2 \cdot \exp\left[-\frac{\mathrm{v_F} b \phi^{3/2} \zeta_{\mathrm{C}}}{V_{\mathrm{m}}}\right]. \tag{37}$$

In Fowler–Nordheim-type natural semi-logarithmic coordinates, this becomes an expression for the natural logarithm $L_{\mathrm{FN}}(V_{\mathrm{m}}^{-1}) \equiv \ln\{I_{\mathrm{m}}/V_{\mathrm{m}}^2\}$:

$$L_{\mathrm{FN}}(V_{\mathrm{m}}^{-1}) = \ln\{A_{\mathrm{fC}}^{\mathrm{SN}} a \phi^{-1} \zeta_{\mathrm{C}}^{-2}\} - \frac{\mathrm{v_F} b \phi^{3/2} \zeta_{\mathrm{C}}}{V_{\mathrm{m}}}. \tag{38}$$

It has been well established since the 1950s [40] that these theoretical FN plots made using 1956 MG theory are very slightly curved, though many experimentalists seem not to realize this. Since 1956 MG FE theory is known to be "better physics" than 1928/29 FN FE theory, expectation is that, in FN plots, experimental data plots would (in the absence of data-noise) lie on a slightly curved line. One cannot, of course, change the nature of the real world by writing down a simplified equation that ignores this curvature effect—as is often done in modern FE technological papers. This is why the data-analysis procedures used in most modern FE technology papers do not yield "best-current-practice" numerical results.

In the *tangent method* of interpreting FE data-analysis plots, the slope $S_{\mathrm{FN}}^{\mathrm{fit}}$ of the straight line fitted to the data-plot points is modelled as a tangent to the theoretical plot made in FN coordinates. For the current–voltage type FN plot, the slope $S_{\mathrm{FN,IV}}^{\mathrm{tan}}$ is given by (using $\bar{V}_{\mathrm{m}}$ to denote $V_{\mathrm{m}}^{-1}$)

$$S_{\mathrm{FN,IV}}^{\mathrm{tan}} = \frac{\mathrm{d}L_{\mathrm{FN}}}{\mathrm{d}\bar{V}_{\mathrm{m}}} = -(b\phi^{3/2}\zeta_{\mathrm{C}}) \cdot \frac{\mathrm{d}(\mathrm{v_F}\bar{V}_{\mathrm{m}})}{\mathrm{d}\bar{V}_{\mathrm{m}}} = -(b\phi^{3/2}\zeta_{\mathrm{C}}) \cdot \left[\mathrm{v_F} + \bar{V}_{\mathrm{m}}\frac{\mathrm{d}\mathrm{v_F}}{\mathrm{d}\bar{V}_{\mathrm{m}}}\right]. \tag{39}$$

where it has been assumed that there is no significant voltage-dependence in $A_{\mathrm{fC}}^{\mathrm{SN}}$, $\phi$ or $\zeta_{\mathrm{C}}$.

Now $\mathrm{d}\bar{V}_{\mathrm{m}} = \mathrm{d}(V_{\mathrm{m}}^{-1}) = -V_{\mathrm{m}}^{-2}\mathrm{d}V_{\mathrm{m}}$, and $V_{\mathrm{m}} = f_{\mathrm{C}} V_{\mathrm{mR}}$, and $\zeta_{\mathrm{C}}$ is constant for an electronically ideal system, and we can replace $\mathrm{v_F}$ by a symbol that shows its explicit dependence on $f_{\mathrm{C}}$, all of which yields

$$S_{\mathrm{FN,IV}}^{\mathrm{tan}}(f_{\mathrm{C}}) = -b\phi^{3/2}\zeta_{\mathrm{C}} \cdot \left[\mathrm{v}(f_{\mathrm{C}}) - f_{\mathrm{C}}\frac{\mathrm{dv}}{\mathrm{d}f_{\mathrm{C}}}\right] = -\mathrm{s}(f_{\mathrm{C}}) \cdot b\phi^{3/2}\zeta_{\mathrm{C}}, \tag{40}$$

where $\mathrm{s}(f_{\mathrm{C}})$ is known as the *slope correction function* and is an appropriate particular version of a FE special mathematical function $\mathrm{s}(x) \equiv \mathrm{v} - x\mathrm{dv}/\mathrm{d}x$.

We now suppose that there is a value of $V_{\mathrm{m}}$ and a corresponding value $f_{\mathrm{C}} = f_{\mathrm{t}}$ at which, in the FN plot, the fitted straight line is parallel to the tangent to the theoretical plot. This value $f_{\mathrm{t}}$ is termed the *fitting value*, and we define $\mathrm{s_t} = \mathrm{s}(f_{\mathrm{t}})$. On identifying $S_{\mathrm{FN}}^{\mathrm{fit}}$ with $S_{\mathrm{FN,IV}}^{\mathrm{tan}}$, we obtain the extracted VCL as

$$\zeta_{\mathrm{C}}^{\mathrm{extr}} = \frac{|S_{\mathrm{FN}}^{\mathrm{fit}}|}{\mathrm{s_t} b\phi^{3/2}}. \tag{41}$$

The slope correction function is an exactly known, weakly varying, function. Initially, the values of $f_{\mathrm{t}}$ and $\mathrm{s_t}$ are not exactly known, but it is usually adequate to make the approximation $\mathrm{s_t} = 0.95$. More precise estimates of $f_{\mathrm{t}}$ and $\mathrm{s_t}$ can in principle be obtained by iteration [41], but usually this is not worthwhile. The value $\mathrm{s_t} = 0.95$ is used in the webtool.

For the 1953 discussion of "s", using the Nordheim parameter $y$, see the letter of Burgess, Kroemer and Houston [40]. For a "21st Century" mathematical discussion, using the scaled field $f_{\mathrm{C}}$, see [42]; there is a modern tabulation of values of "s" in [43].

*Fowler–Nordheim plots: extraction of area-like parameters.* For the extraction of formal emission area, we have preferred to use the *extraction parameter approach*, in which a formula for formal emission area is written in the form

$$\{A_{\mathrm{fC}}^{\mathrm{SN}}\}^{\mathrm{extr}} = \Lambda_{\mathrm{FN}}^{\mathrm{SN}} R_{\mathrm{FN}}^{\mathrm{fit}} (S_{\mathrm{FN}}^{\mathrm{fit}})^2, \tag{42}$$

where $\Lambda_{\mathrm{FN}}^{\mathrm{SN}}$ is a so-called *extraction parameter*. The subscript position indicates the type of data-analysis plot being analysed; the superscript position indicates the type of tunnelling barrier being assumed in the theoretical analysis.

The value of the extraction parameter depends both on local work function and on the fitting value $f_{\mathrm{t}}$, and is given by the formula

$$\Lambda_{\mathrm{FN}}^{\mathrm{SN}}(\phi, \mathrm{r_t}) = \frac{1}{[\mathrm{r_t} \mathrm{s_t}^2 a b^2 \phi^2]}, \tag{43}$$

where $\mathrm{r_t}$ is the fitting value of the *2012 intercept correction factor* discussed in [42] and denoted in the present paper by $\mathrm{r}(\phi, f_{\mathrm{C}})$.

The parameter $r_{\mathrm{t}}$ is in fact a relatively sensitive function both of $f_{\mathrm{t}}$ and of the local work-function $\phi$. The $\phi$-dependence is illustrated in Fig. 6. If precise values are needed with respect to $f_{\mathrm{t}}$, then iteration as described above can be used. (But with modern data, as opposed to historical data, it will be better to use a Murphy–Good plot, as described below.)

Because a FN plot involves fitting a straight line to data points that are known to lie on a curve, it follows that in principle a "chord correction" ought to be applied to area extraction by the tangent method. In reality the correction is very small (and the need goes away if a MG plot is used).

Apart from the tangent method, there are FE data-analysis methods based on fitting chords to FN plots and others based on "linearizing the emission equation”. However, these approaches also have limitations. Further, each different extraction method will or may return a slightly different extracted area value from a given single set of input data, thereby making scientific comparisons of reported values problematic. One objective of designing the Murphy–Good plot was to make problems of this general kind obsolete, and specify a data-analysis approach that (within the more general limitations of SPME methodology) returns a result that depends only on the assumed uniform value of local work-function.

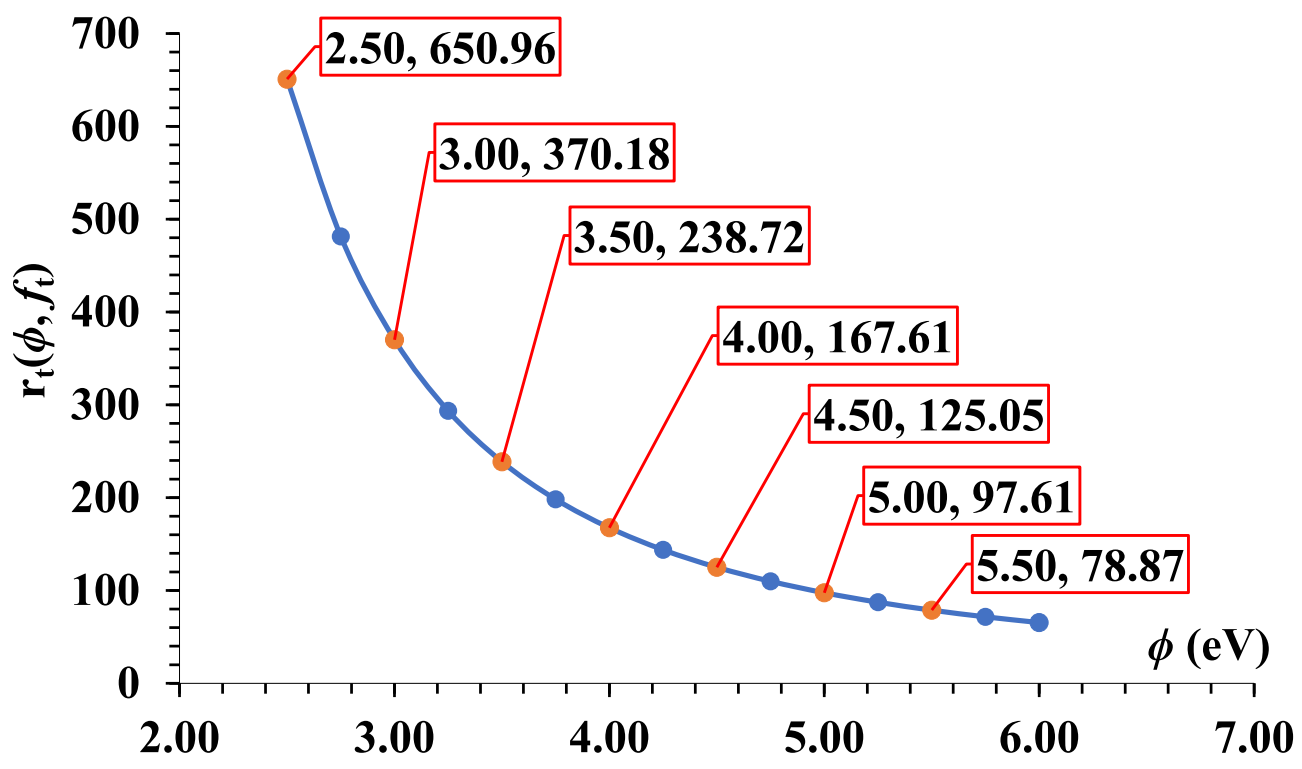

**Fig. 6.** In the context of the tangent method, to show how the vertical-axis intercept correction factor $r_t$ depends on the local work function $\phi$ of the emitting surface, for the specific fitting value $f_t = 0.2815163$. This value corresponds precisely (to 7 sig. fig.) to the assumption that $s(f_t) = 0.95$ exactly.

*Millikan–Lauritsen (ML) plots*. Extraction of VCL and related values from the slope of an ML plot can be done by finding the equivalent slope for a FN plot, using Eq. (34), and using the theory above for a FN plot. At present, there is no standard methodology for extracting values of formal emission area.

*Murphy–Good (MG) plots*. The theory and merits of MG plots have recently been discussed elsewhere [7], and are not examined in detail here. The theory derives from the expanded scaled form of MG FE theory and uses the fact that, in Eq. (16) for the kernel current density (for the SN barrier), the characteristic scaled field appears as an non-integral power $\kappa^{SN}$ given by

$$\kappa^{SN} = 2 - \eta/6. \tag{44}$$

Hence, by making a data-analysis plot with index $n = 2 - \eta/6$, a plot should be obtained that is “very nearly straight”. The reservation “very nearly” arises because the so-called “simple good approximation” is not exact; however, the linearity is significantly better than that of a FN plot. Formulae for extracted characterization parameters are

$$\zeta_C^{extr} = \frac{|S_{MG}^{fit}|}{b\phi^{3/2}}. \tag{45}$$

$$\{A_{fC}^{SN}\}^{extr} = \Lambda_{MG}^{SN} R_{MG}^{fit} (|S_{MG}^{fit}|)^{\kappa}, \tag{46}$$

$$\Lambda_{MG}^{SN}(\phi) = \frac{1}{[\exp(\eta) \cdot \eta^{-\eta/6} \cdot ab^2\phi^2]}, \tag{47}$$

where $S_{MG}^{fit}$ and $R_{MG}^{fit}$ are the slope and intercept of a straight line fitted to a MG plot. A table of values of the extraction parameter is provided in [7]. There is only a moderate dependence on the assumed local-work-function value.

## 3. Webtool overview

The webtool is built on the mathematical assumptions of FE theory, with JavaScript used as a programming language. The webtool is self-contained in the sense that the specified inputs are used to calculate all relevant working parameters, as well as to generate Orthodoxy-Test results and (where relevant) emitter characterization-parameter values. Inputs are validated roughly for incorrect and/or illogical values, like empty cells, zero values or negative values of $\phi$.

Required inputs are set out in the flowchart shown as Fig. 7. Optional inputs are the system geometrical parameters $d_M$ and $A_M$. The webtool then operates as follows. The slope of the fitted line that passes through the two range limits is calculated. The vertical ($\bar{X} = 0$) axis intercept is then calculated using the coordinates of the left-hand limit and the generated slope value. The input local work-function is then used to extract the characteristic-scaled-field values ($f_C$-values) that correspond to the two range-limits. Next, the input local work function is rounded to the nearest first decimal value listed in Table 2. This rounded work-function value is then used to determine the relevant orthodoxy-test zone boundaries. The test can then be applied to the extracted range of scaled-field values, and the result reported. Where relevant, the webtool also reports precise values of the $\phi$-related characterization parameters $\eta(\phi)$ and $\theta(\phi)$, and the pre-exponential parameter $\kappa^{SN}$.

The requirements for applying the FE Orthodoxy Test (by itself) to any of the data-analysis plots are simple, since the only input-data requirements are the assumed local work function of the emitter surface and the coordinates of the two range-limits of the data-analysis plot under test. Further, the test is known to be robust, in that the obtained results do not depend on the units of measurements used for the horizontal axis and for the arguments of the vertical-axis logarithms. Thus, if the only objective is to apply the Orthodoxy Test, then the user is not obliged to work with plots made using only the units A, V and m.

However, if it is required to extract values of emitter characterization parameters, then it is NECESSARY to use only data-input values involving the units A, V, and m, NOT submultiples of these units. This applies to values of $\bar{X}$, to the arguments of logarithms, and to the values of the geometrical parameters $d_M$ and $A_M$.

The webtool will display values of emitter characterization parameters only if the Orthodoxy Test is passed. Depending on the nature of the data-input format, the “extracted parameters” in Table 2 will automatically be displayed. The “derived parameters” will be displayed if values of $d_M$ and $A_M$ have already been entered, or are then entered.

In theory, if the Orthodoxy Test is failed then it may sometimes be possible to use the procedure of *phenomenological adjustment* [44] to convert a spurious extracted VCL or FEF value into a rough estimate of the true value, but this option is not incorporated in the current version of the webtool.

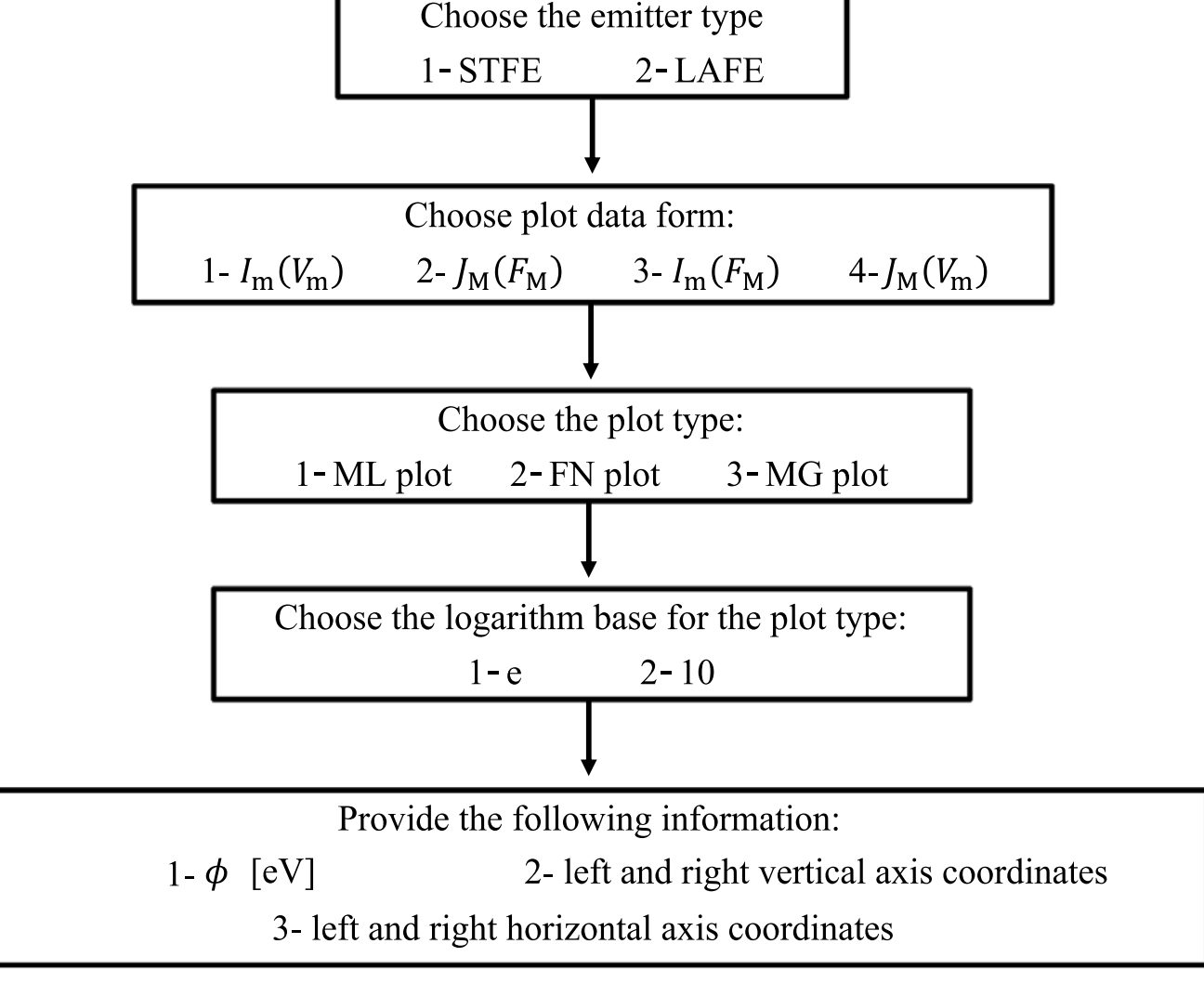

**Fig. 7.** Information inputs needed in order to use the webtool (in its present form) to apply the Orthodoxy Test.

**Table 3**
Simulated data values for input into the webtool, in order to compare extraction performances. Data are generated using EMG theory and the input values $\phi = 4.50$ eV, $\zeta_C = 200$ nm and $A_f^{SN} = 100$ nm$^2$. All quantities are measured using the SI units A, V and m.

| $f_C \rightarrow$ | 0.30 | 0.43 | 0.55 | 0.77 |
| --- | --- | --- | --- | --- |
| $V_m$ | 844 | 1209 | 1547 | 2166 |
| $F_M$ | $8.44\times10^5$ | $1.21\times10^6$ | $1.55\times10^6$ | $217\times10^6$ |
| $I_m$ | $3.20\times10^{-8}$ | $5.24\times10^{-6}$ | $7.39\times10^{-5}$ | $1.23\times10^{-3}$ |
| $J_M$ | $3.20\times10^{-4}$ | $5.24\times10^{-2}$ | 0.74 | 12.33 |
| $1/V_m$ | $1.19\times10^{-3}$ | $8.27\times10^{-4}$ | $6.46\times10^{-4}$ | $4.62\times10^{-4}$ |
| $1/F_M$ | $1.19\times10^{-6}$ | $8.27\times10^{-7}$ | $6.46\times10^{-7}$ | $4.62\times10^{-7}$ |
| $\ln\{I_m\}$ | −17.26 | −12.16 | −9.51 | −6.70 |
| $\ln\{I_m/V_m^2\}$ | −30.73 | −26.36 | −24.20 | −22.06 |
| $\ln\{I_m/V_m^\kappa\}$ | −25.53 | −20.87 | −18.53 | −16.12 |
| $\ln\{I_m/F_M^2\}$ | −44.55 | −40.17 | −38.02 | −35.87 |
| $\ln\{I_m/F_M^\kappa\}$ | −34.00 | −29.35 | −27.00 | −24.60 |
| $\ln\{J_M\}$ | −8.05 | −2.95 | −0.30 | 2.51 |
| $\ln\{J_M/F_M^2\}$ | −35.34 | −30.96 | −28.81 | −26.66 |
| $\ln\{J_M/F_M^\kappa\}$ | −24.79 | −20.14 | −17.79 | −15.39 |
| $\ln\{J_M/V_m^2\}$ | −21.52 | −17.15 | −14.99 | −12.85 |
| $\ln\{J_M/V_m^\kappa\}$ | −16.32 | −11.66 | −9.32 | −6.91 |

## 4. Simulations of webtool use

### 4.1. Simulation input data

This Section describes our testing procedure for the functionality and software engineering of the webtool. Assumed emitter and system characterization parameters are: $A_M = 100$ mm$^2$; $d_M = 1$ mm; and, as previously mentioned: $A_{fC}^{SN} = 100$ nm$^2$, $\zeta_C = 200$ nm, and $\phi = 4.50$ eV. These yield $\gamma_{MC} = 5000$ and $\alpha_{fC}^{SN} = 10^{-12}$. For MG plots, the value $\phi = 4.50$ eV yields the precise value $\kappa^{SN} = 1.227192$. Three scaled-field ranges have been chosen, as described below, in order to allow coverage of to all possible orthodoxy-test outcomes . The corresponding $I_m(V_m)$ characteristics were shown in Fig. 4, for the whole scaled-field range.

The simulations have been carried out for all four data-input formats and all three data-analysis plot types. Thus, this simulation exercise covers all the variations commonly used in the literature.

The total scaled-field range used has been $0.30 \leqslant f_C \leqslant 0.77$. This range has been also used to generate Fig. 4. Within this overall range, three sub-ranges have been considered: (0.30–0.43), (0.30–0.55), and (0.30–0.77). The resulting calculated data needed for input into the webtool are presented in Table 3.

### 4.2. Webtool output data

Tables 4 and 5 present output data from the webtool. Table 4 shows the results of extracting the range-limiting $f_C$-values, for the various plot types and data forms, together with the errors as compared with the values input into the simulations. Three main conclusions can be drawn from these results.

(1) For each data-plot type, the errors for the $J_M(V_m)$ cases are the same as those for the $I_m(V_m)$ cases (e.g., lines "10" and "1"), and the errors for $J_M(F_M)$ cases are the same as those for the $I_M(F_M)$ (e.g., lines "7" and "4"). This is presumably because, for any given emitter, the ratio $I_m/J_M[= A_M]$ is a system related constant. Thus, any future comparisons do not need to investigate the data forms involving the macroscopic current density $J_M$.

(2) In the simulations, when extracting values of scaled field and the characterization parameters, it was found that using the original input current–voltage characteristics gave more accurate results than when the measured voltage was converted to the related macroscopic field.

(3) Both when extracting values of characteristic scaled field, and when extracting values of the voltage conversion length and (where possible) the formal emission area, the use of a current–voltage Murphy–Good plot is seen to involve *smaller* methodology-errors than use of either a FN plot or a ML plot (data-noise errors are not part of the present discussion). This strongly suggests that it would be advantageous for those interested in analysing FE current–voltage data to move to the use of Murphy–Good plots, rather than FN plots (or ML plots).

As already indicated, the essential reason for the superior performance of the MG plot is that the FN plot is slightly curved, but the MG plot is "very nearly straight". This can be illustrated by using EMG FE theory, and the $I_m(V_m)$ data form, to predict how the slopes of the three types of data plot vary with measured voltage. Using current–voltage simulation results within the orthodoxy-test pass-range of $f_C$ values $0.15 \leqslant f_C \leqslant 0.45$, the total variation for the slope has been investigated. For illustration, results have been obtained by evaluating the slope at several points within the interval $0.31 \leqslant f_C \leqslant 0.43$, At each point the slope is evaluated using two very-close points around the corresponding $f_C$ value. The results are presented in Table 5.

The slope variations from Table 5 demonstrate that (when EMG theory is used in the simulations) then neither a FN plot nor a ML plot is exactly straight, but that an MG plot is "very nearly straight". Thus, MG plots are clearly a better analysis tool (if precise extraction of intercept-related parameters is the goal).

However, it is important to remember that these simulations are carried out within the framework of "smooth planar metal-like emitter (SPME) methodology". If the emitting surface has significant local curvature, then both experimental MG plots and MG plots simulated using curved-emitter emission theory will be slightly curved. Discussion of such effects is outside the scope of this paper.

## 5. Conclusions and future developments

### 5.1. Conclusions

This research has used Extended Murphy–Good (EMG) FE theory to generate simulated FE current–voltage data, and has then used this data to explore the processes of applying the FE Orthodoxy Test and of extracting field emitter characterization parameters. This has been set in the context of summaries of EMG emission theory and of the electronic engineering of FE systems, and of a detailed discussion of why validity checks should be a standard part of FE data analysis. Using the webtool previously developed by one of us (MMA), a thorough comparison has been made of the merits of the three conventional data-plot types (Millikan–Lauritsen, Fowler–Nordheim and Murphy–Good), using all four of the data-input formats that have been used (either measured voltage or apparent macroscopic field as the independent variable, and either measured current or macroscopic current density as the dependent variable). In particular, we have compared plot slope values, extracted scaled-field values, and extracted values of voltage conversion length and formal emission area.

This detailed discussion of the electronic engineering of FE systems has re-confirmed that significantly the best data-input method is to use the raw experimental current–voltage data, with voltages measured in volts and currents in amperes (not in submultiples of these units).

The general consistency of the results obtained demonstrates that the webtool is functioning effectively. The analysis has also confirmed that use of a Murphy–Good plot is a methodology that has a parameter-extraction precision superior to that of either a FN plot or a ML plot, particularly in respect of the extraction of formal emission areas. This confirmation is particularly important in the context of developing better FE science, where it is now known that important differences as between different theories can manifest themselves in the formal emission area, rather than in the FE equation exponent.

**Table 4**
Results of extracting scaled field values and characterization parameters, using the webtool. The characterization parameters are compared to the original input values $\zeta_C = 200$ nm and $A_{fC}^{SN} = 100$ nm$^2$. All the quantities are measured using the SI units A, V and m. N/A indicates that results are not available because they are not valid because they failed the orthodoxy test.

| Data Form | Num. | Slope Np $\cdot[X]$ | Tested $f_C$-Range | Extracted $f_C$ values | | | | $A_{fC}^{SN}$ nm$^2$ | Error | $\zeta_C$ nm | Error | $\alpha_{fC}^{SN}$ | $\gamma_C$ |
|---|---|---|---|---|---|---|---|---|---|---|---|---|---|
| | | | | $f_{low}^{extr}$ | Error | $f_{up}^{extr}$ | Error | | | | | | |
| MG plots | | | | | | | | | | | | | |
| $I_m(V_m)$ | 1 | −12944 | 0.3–0.43 | 0.3010 | 0.34% | 0.4316 | 0.37% | 96.92 | −3.08% | 198.51 | −0.74% | 0.96 | 5037 |
| | 2 | −12963 | 0.3–0.55 | 0.3006 | 0.20% | 0.5503 | 0.06% | 99.25 | −0.75% | 198.80 | −0.60% | 0.99 | 5030 |
| | 3 | −12890 | 0.3–0.77 | 0.3023 | 0.76% | 0.7820 | 1.56% | N/A | N/A | N/A | N/A | N/A | N/A |
| $I_m(F_M)$ | 4 | $-1.28\times10^{7}$ | 0.3–0.43 | 0.3042 | 1.39% | 0.4377 | 1.79% | 82.11 | −17.89% | 196.45 | −1.78% | 0.82 | 5090 |
| | 5 | $-1.29\times10^{7}$ | 0.3–0.55 | 0.3021 | 0.69% | 0.5564 | 1.17% | 88.43 | −11.57% | 197.33 | −1.33% | 0.88 | 5068 |
| | 6 | $-1.29\times10^{7}$ | 0.3–0.77 | 0.3021 | 0.69% | 0.7780 | 1.04% | N/A | N/A | N/A | N/A | N/A | N/A |
| $J_M(F_M)$ | 7 | $-1.28\times10^{7}$ | 0.3–0.43 | 0.3042 | 1.39% | 0.4377 | 1.79% | 82.08 | −17.92% | 196.45 | −1.78% | 0.82 | 5090 |
| | 8 | $-1.29\times10^{7}$ | 0.3–0.55 | 0.3021 | 0.69% | 0.5564 | 1.17% | 88.41 | −11.59% | 197.33 | −1.33% | 0.88 | 5068 |
| | 9 | $-1.29\times10^{7}$ | 0.3–0.77 | 0.3021 | 0.69% | 0.7780 | 1.04% | N/A | N/A | N/A | N/A | N/A | N/A |
| $J_M(V_m)$ | 10 | −12944 | 0.3–0.43 | 0.3010 | 0.34% | 0.4316 | 0.37% | 96.88 | −3.12% | 198.51 | −0.74% | 0.96 | 5037 |
| | 11 | −12963 | 0.3–0.55 | 0.3006 | 0.20% | 0.5503 | 0.06% | 99.22 | −0.78% | 198.80 | −0.60% | 0.99 | 5030 |
| | 12 | −12890 | 0.3–0.77 | 0.3023 | 0.76% | 0.7820 | 1.56% | N/A | N/A | N/A | N/A | N/A | N/A |
| FN plots | | | | | | | | | | | | | |
| $I_m(V_m)$ | 13 | −12139 | 0.3–0.43 | 0.3049 | 1.65% | 0.4372 | 1.68% | 75.88 | −24.12% | 195.96 | −2.02% | 0.75 | 5103 |
| | 14 | −12092 | 0.3–0.55 | 0.3061 | 2.04% | 0.5604 | 1.89% | 71.27 | −28.73% | 195.21 | −2.39% | 0.71 | 5123 |
| | 15 | −11877 | 0.3–0.77 | 0.3117 | 3.89% | 0.8063 | 4.71% | N/A | N/A | N/A | N/A | N/A | N/A |
| $I_m(F_M)$ | 16 | $-1.21\times10^{7}$ | 0.3–0.43 | 0.3059 | 1.97% | 0.4402 | 2.37% | 68.45 | −31.55% | 194.78 | −2.61% | 0.68 | 5134 |
| | 17 | $-1.20\times10^{7}$ | 0.3–0.55 | 0.3085 | 2.82% | 0.5682 | 3.32% | 62.89 | −37.11% | 193.77 | −3.11% | 0.63 | 5161 |
| | 18 | $-1.19\times10^{7}$ | 0.3–0.77 | 0.3111 | 3.69% | 0.8012 | 4.06% | N/A | N/A | N/A | N/A | N/A | N/A |
| $J_M(F_M)$ | 19 | $-1.21\times10^{7}$ | 0.3–0.43 | 0.3059 | 1.97% | 0.4402 | 2.37% | 68.24 | −31.76% | 194.78 | −2.61% | 0.68 | 5134 |
| | 20 | $-1.20\times10^{7}$ | 0.3–0.55 | 0.3085 | 2.82% | 0.5682 | 3.32% | 62.87 | −37.13% | 193.77 | −3.11% | 0.63 | 5161 |
| | 21 | $-1.20\times10^{7}$ | 0.3–0.77 | 0.3111 | 3.69% | 0.8012 | 4.06% | N/A | N/A | N/A | N/A | N/A | N/A |
| $J_M(V_m)$ | 22 | −12139 | 0.3–0.43 | 0.3049 | 1.65% | 0.4372 | 1.68% | 75.86 | −24.14% | 195.96 | −2.02% | 0.76 | 5103 |
| | 23 | −12092 | 0.3–0.55 | 0.3061 | 2.04% | 0.5604 | 1.89% | 71.25 | −28.75% | 195.21 | −2.39% | 0.71 | 5123 |
| | 24 | −11877 | 0.3–0.77 | 0.3117 | 3.89% | 0.8063 | 4.71% | N/A | N/A | N/A | N/A | N/A | N/A |
| ML plots | | | | | | | | | | | | | |
| $I_m(V_m)$ | 25 | −12186 | 0.3–0.43 | 0.3038 | 1.25% | 0.4355 | 1.28% | N/A | N/A | 196.72 | −1.64% | N/A | 5083 |
| | 26 | −12178 | 0.3–0.55 | 0.3040 | 1.32% | 0.5565 | 1.18% | N/A | N/A | 196.59 | −1.70% | N/A | 5087 |
| | 27 | −12042 | 0.3–0.77 | 0.3074 | 2.47% | 0.7953 | 3.28% | N/A | N/A | N/A | N/A | N/A | N/A |
| $I_m(F_M)$ | 28 | $-1.21\times10^{7}$ | 0.3–0.43 | 0.3059 | 1.97% | 0.4402 | 2.37% | N/A | N/A | 194.79 | −2.60% | N/A | 5134 |
| | 29 | $-1.21\times10^{7}$ | 0.3–0.55 | 0.3059 | 1.97% | 0.5635 | 2.46% | N/A | N/A | 194.81 | −2.59% | N/A | 5134 |
| | 30 | $-1.21\times10^{7}$ | 0.3–0.77 | 0.3059 | 1.97% | 0.7880 | 2.34% | N/A | N/A | N/A | N/A | N/A | N/A |
| $J_M(F_M)$ | 31 | $-1.21\times10^{7}$ | 0.3–0.43 | 0.3059 | 1.97% | 0.4402 | 2.37% | N/A | N/A | 194.79 | −2.60% | N/A | 5134 |
| | 32 | $-1.21\times10^{7}$ | 0.3–0.55 | 0.3059 | 1.97% | 0.5635 | 2.46% | N/A | N/A | 194.81 | −2.59% | N/A | 5134 |
| | 33 | $-1.21\times10^{7}$ | 0.3–0.77 | 0.3059 | 1.97% | 0.7880 | 2.34% | N/A | N/A | N/A | N/A | N/A | N/A |
| $J_M(V_m)$ | 34 | −12186 | 0.3–0.43 | 0.3038 | 1.25% | 0.4355 | 1.28% | N/A | N/A | 196.72 | −1.64% | N/A | 5083 |
| | 35 | −12178 | 0.3–0.55 | 0.3040 | 1.32% | 0.5565 | 1.18% | N/A | N/A | 196.59 | −1.70% | N/A | 5087 |
| | 36 | −12042 | 0.3–0.77 | 0.3074 | 2.47% | 0.7953 | 3.28% | N/A | N/A | N/A | N/A | N/A | N/A |

**Table 5**
Total variation of the simulated slope values that are obtained from each of the current–voltage data-analysis plot types, over the range $0.31 \leqslant f_C \leqslant 0.43$. Note that the neper (Np) is the "amplitude" unit of difference in natural logarithms, and is a marker that natural logarithms are being used.

| $f_C$ | $S_{ML}$ (Np · V) | $S_{FN}$ (Np · V) | $S_{MG}$ (Np · V) |
|---|---|---|---|
| 0.32 | −14104 | −12304 | −12999.21 |
| 0.34 | −14172 | −12259 | −12998.37 |
| 0.36 | −14240 | −12215 | −12997.69 |
| 0.38 | −14309 | −12171 | −12997.16 |
| 0.40 | −14377 | −12127 | −12996.77 |
| 0.42 | −14446 | −12084 | −12996.53 |
| Total variation | 342 | 220 | 2.7 |

### 5.2. Future developments — theoretical

As already indicated, the next major development stage in FE current–voltage data analysis will be to use emission theory for curved emitters. (Whether existing curved-emitter theory is exactly correct from the point of view of quantum mechanics is another matter — for example it is still an "atom-free" theory; however, existing curved-emitter FE theory is "better physics" than existing planar-emitter FE theory.) This will bring another emitter characterization parameter – the local surface radius of curvature – into the discussion.

When FE experimentalists are confronted with TWO incompletely known parameters (work function and radius of curvature), plotting methods of the types discussed here look as though they will become significantly less effective, even if implemented correctly. Methodologies based on multi-variable numerical regression look a better choice (for example, see [34]). However, our view is that numerical methodologies of this kind should first be tried out on the planar-emission situation, using the work-function value as a third unknown variable (in addition to the VCL and formal emission area). The details of how data noise will affect numerical results of this kind (especially statistical error limits) are far from obvious, and there is scope for relevant simulations.

### 5.3. Future developments — webtool applications

For most of the authors, our particular interests are in improved electron microscope (EM) sources, including hybrid-design sources involving dielectric layers on metal needle-like substrates (which may be able to operate in poorer vacuum conditions than bare metal needles). We expect the webtool to be useful in these applications, but we also expect the webtool to be useful more generally in other research and development activities relating to FE based sources and devices.

As specific examples within the context of electron sources, the webtool has been used to analyse results from tungsten needle-like emitters [45] and from LAFEs of polymer graphite flakes [46]. We plan that the webtool should be a first step in a larger project that aims to monitor and control the behaviour of electron sources during operation, not only in FE experiments, but also in related scientific and industrial instruments, for example field emission scanning electron microscopes.

The webtool is still under development, but in its current state may be accessed via [47]. This paper has described the checks we have made on basic aspects of its software engineering, but (thanks to comments made by an anonymous reviewer) we have set this in the context of a much more general discussion of the underlying theory and of why validity checks on measured current–voltage data (and on converted versions of this data) should be a standard part of FE data-analysis procedures.

Finally, we emphasize again the following three points. (1) Best scientific and engineering practice is to make plots using the raw experimental data (measured voltages in V and measured currents in A). (2) For orthodoxly behaving FE systems, the Murphy–Good plot provides better accuracy than the FN plot, particularly as regards the extraction of formal emission area. (3) ALWAYS apply a validity check (preferably the Orthodoxy Test, and preferably using a MG plot) before extracting values for characterization parameters. We believe that our webtool (and any future extensions) can help the user do these things, and that using it will provide an easy and convenient way of interpreting FE current–voltage characteristics, albeit within the framework of "the smooth planar metal-like emitter" methodology and "21st Century planar FE theory".

## Declaration of competing interest

The authors declare that they have no known competing financial interests or personal relationships that could have appeared to influence the work reported in this paper.

## Data availability

The data that supports the simulation findings of this study are available within the article.

## Acknowledgements

The research described in this paper was financially supported by the Ministry of the Interior of the Czech Republic (project. No. VI20192022147). We also acknowledge the Czech Academy of Sciences (RVO:68081731).

The authors would like to acknowledge the support of Mu'tah University through the research project #452/2021.

## Appendix. Definition of a dimensionless PPP-geometry field enhancement factor (PFEF)

Fig. 8 illustrates the definition of the type of field enhancement factor most commonly used in the theory given in FE literature, called here a "characteristic true plate-field enhancement factor". The interpretation of the labels is as follows: "plate-field" indicates that the type of macroscopic field being considered is that between two parallel planar plates; "true" indicates that the voltage used in the definition is that between the plates (this voltage may or may not be equal to the measured voltage); "characteristic" means the local field is taken at some location that characterizes the emitter behaviour (the location where the local current density is highest is in principle usually best: in modelling this usually coincides with the emitter apex).

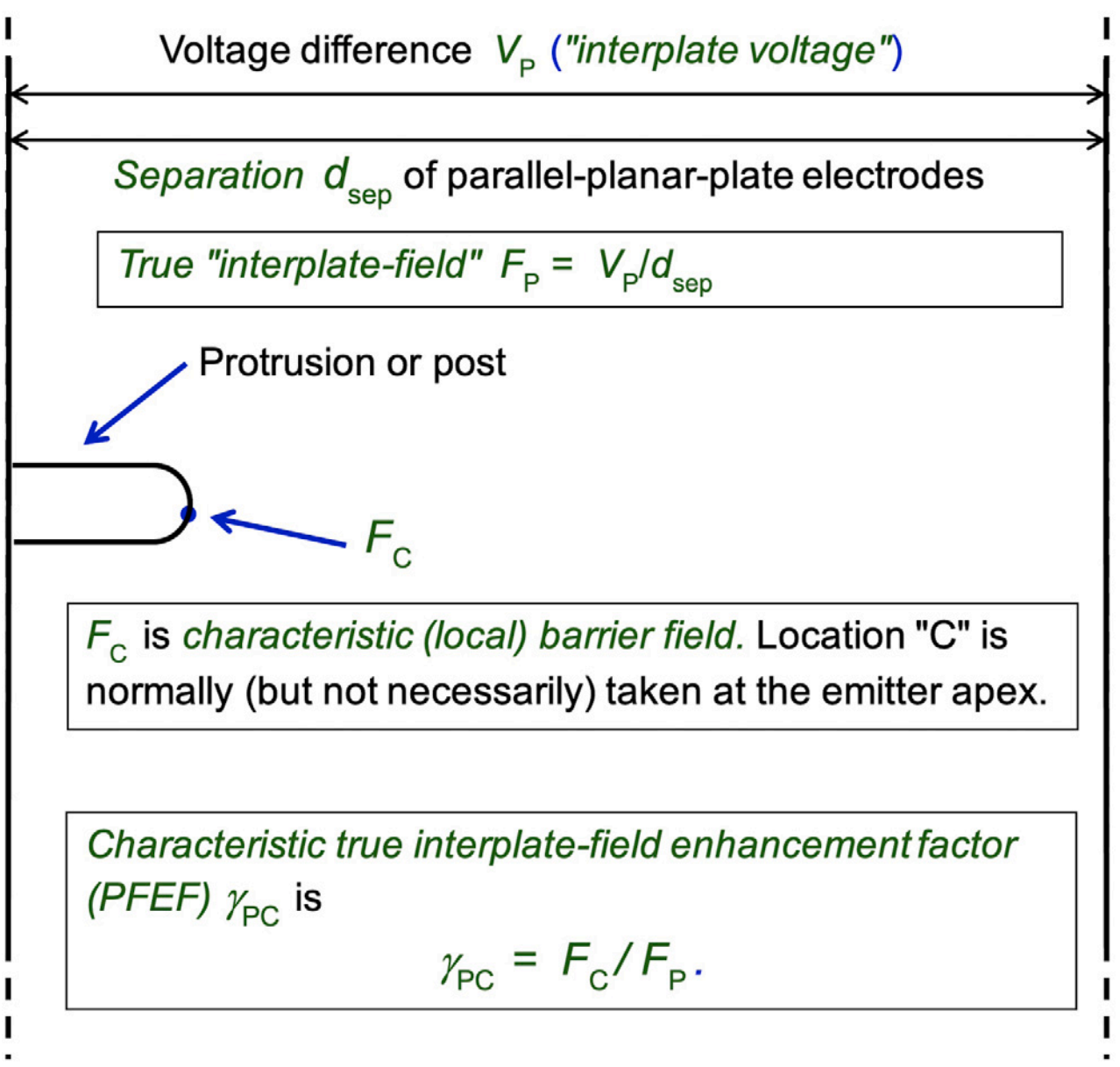

**Fig. 8.** To illustrate the definition of the "plate-field enhancement factor (PFEF)" used in parallel-planar-plate (PPP) emission geometry.